\documentclass{article}

\usepackage{fullpage}

\newcommand{\egdef}{\stackrel{\mbox{\begin{tiny}def\end{tiny}}}{=}}
\newcommand{\tuple}[1]{\langle #1 \rangle}              
\newcommand{\tq}{\mid}                                  
\newcommand{\sat}{\models}                              
\newcommand{\size}[1]{\mid\! #1 \!\mid}

\newcommand{\et}{\wedge}

\newcommand{\union}{\cup}

\newcommand{\fleche}{\rightarrow}                       
\newcommand{\flsup}[1]{\stackrel{#1}{\rightarrow}}

\newcommand{\theoname}[1]{{\bf (#1)}}

\newcommand{\ou}{\ensuremath{\vee} }

\newcommand{\alws}{\ensuremath{\Box}}
\newcommand{\mynext}{\ensuremath{\bigcirc}}
\newcommand{\event}{\ensuremath{\diamondsuit}}
\newcommand{\until}{\ensuremath{\mathcal{U}}}

\newcommand{\Impl}{\ensuremath{\Rightarrow}}

\newcommand{\BDef}{\ensuremath{\hat{=}}}

\newcommand{\Or} {\vee}

\newcommand{\alw} {\Box}
\newcommand{\nxt} {\bigcirc}
\newcommand{\evtly} {\diamondsuit}

\newcommand{\domaine}{D}
\newcommand{\ensvar}{V} 
\newcommand{\propat}{\ensuremath{{ap}}}
\newcommand{\enspropat}{\ensuremath{AP}}

\newcommand{\TS}{\ensuremath{{TS}}}
\newcommand{\FTS}{\ensuremath{{F\TS}}}

\newcommand{\etat}{s} 
\newcommand{\s}{\etat} 
\newcommand{\ensetat}{S} 
\newcommand{\init}{\ensetat_{0}}
\newcommand{\act}{a} 
\newcommand{\ensact}{{Act}}
\newcommand{\enstrans}{\rightarrow} 
\newcommand{\fdecor}{L}
\newcommand{\enstranseq}{\mathcal{F}} 

\newcommand{\transe}[3]{\ensuremath{#1 \stackrel{#2}{\to} #3}}

\newcommand{\TSnorm}{\tuple{\init,\ensetat,\linebreak[1]\ensact,\linebreak[1]\enstrans,\linebreak[1]\fdecor}}

\newcommand{\FTSnorm}{\tuple{\init,\ensetat,\ensact,\enstrans,\fdecor,\enstranseq}}

\newcommand{\nsat}{\not\sat} 

\newcommand{\inv}{I}

\newcommand{\chem}{\sigma} 
\newcommand{\enschem}[1]{\Sigma_{#1}}

\newcommand{\AB}{\mathcal{B}} 
\newcommand{\etatAB}{q} 
\newcommand{\q}{\etatAB} 
\newcommand{\initAB}{\etatAB_0} 
\newcommand{\ensetatAB}{Q} 
\newcommand{\ensetiqAB}{SP_V} 
\newcommand{\enstransAB}{\rightarrow} 
\newcommand{\accept}{A} 
\newcommand{\ABmod}{\AB_{{mod}}} 
\newcommand{\chemAB}{\pi} 
\newcommand{\ABnorm}{\tuple{\initAB,\ensetatAB,\ensetiqAB,\enstransAB_{\AB},\accept}}

\newcommand{\invcol}{\inv_{12}}

\newcommand{\module}{M} 

\newcommand{\ensmodule}{\mathbb{M}}

\newcommand{\TSnormind}[1]{\tuple{\ensetat_{0_{#1}},\ensetat_{#1},\ensact_{#1},
\enstrans_{#1},\fdecor_{#1}}}

\newcommand{\FTSnormind}[1]{\tuple{\ensetat_{0_{#1}},\ensetat_{#1},\ensact_{#1},
\enstrans_{#1},\fdecor_{#1}, \enstranseq_{#1}}}

\newcommand{\FTStaunormind}[1]{\tuple{\ensetat_{0_{#1}},\ensetat_{#1},{\ensact_1}_{\tau},
{\enstrans_{#1}}_{\tau},\fdecor_{#1}, \enstranseq_{#1}}}

\newcommand{\eqclass}[1]{EC(#1)}

\usepackage{amssymb}
\usepackage{latexsym}
\usepackage{psfrag}
\usepackage{amsmath, version}
\usepackage{amsfonts}
\usepackage[dvips]{graphicx}
\usepackage{epsfig}
\usepackage{psfig}
\usepackage{xypic}
\usepackage{subfigure}
\usepackage{boxedminipage}
\usepackage{array}

\newtheorem{theorem}{Theorem}[section]

\newtheorem{property}[theorem]{Property}
\newtheorem{lemma}[theorem]{Lemma}
\newtheorem{definition}[theorem]{Definition}
\newtheorem{remark}[theorem]{Remark}

\newenvironment{proof}{\noindent\textbf{Proof.}~}{\hfill$\Box$}

\title{PLTL Partitioned Model Checking \\
for Reactive Systems \\
under Fairness Assumptions}

\author{S.~\textsc{Chouali}, J.~\textsc{Julliand},
P.-A.~\textsc{Masson} and F.~\textsc{Bellegarde} \\ Laboratoire
d'Informatique de l'Universit\'{e} de Franche-Comt\'{e},\\ FRANCE -
LIFC FRE CNRS 2661 }

\date{}

\begin{document}


\maketitle
\begin{center}
\textit{ACM Transactions on Embedded Computing Systems
(TECS)}, 4(2):267--301, May 2005
\end{center}

\begin{abstract}
We are interested in verifying dynamic properties  of finite state
reactive systems under fairness assumptions by model checking. The
systems we want to verify are specified through a top-down
refinement process.

In order to deal with the state explosion problem, we have
proposed in previous works to partition the reachability graph,
and to perform the verification on each part separately. Moreover,
we have defined a class, called $\ABmod$, of dynamic properties
that are \emph{verifiable by parts}, whatever the partition. We
decide if a property $P$ belongs to $\ABmod$ by looking at the
form of the B\"{u}chi automaton that accepts $\neg P$.  However,
when a property $P$ belongs to $\ABmod$, the property $f
\Rightarrow P$, where $f$ is a fairness assumption, does not
necessarily belong to $\ABmod$.

In this paper, we propose to use the refinement process in order
to build the parts on which the verification has to be performed.
We then show that with such a partition, if a property $P$ is
verifiable by parts and if $f$ is the expression of the fairness
assumptions on a system, then the property $f\Impl P$ is still
verifiable by parts.

 This approach is illustrated by its application to
the chip card protocol T=1 using the $B$ engineering design
language.

\ \\
\noindent\textbf{keywords.}~Refinement design, PLTL model checking, fairness assumptions,
out-of-core model checking.

\end{abstract}


\section{Motivations}
This paper is about the verification of dynamic properties of
finite state systems. In our approach, reactive systems are
modeled by transition systems expressed as event systems, for
example in B \cite{abrial96b}, and are specified through a
top-down refinement process.  We propose to express the dynamic
properties (safety, liveness) as formulae of the Propositional
Linear Temporal Logic (PLTL) \cite{Manna&92,Abrial&98}, and to
verify them by model checking \cite{Queille82,Clarke86,Clarke99}.

It is well known that the main drawback of the PLTL model
checking~\cite{Lichtenstein85,Vardi86} is that it cannot handle
very large finite systems.  To avoid the model checking explosion,
many solutions have been proposed, such as partial order
\cite{Katz88,Wolper93}, abstraction
\cite{Cousot77,Clarke94,Dingel&95} and symbolic representation by
BDD \cite{Burch90,McMillan93}.  For a class of PLTL properties, we
propose another solution which is compatible with the previous
ones.

\subsection{Our Propositions}
In order to deal with the problem of the exponential blow up of
the PLTL model checking, we have proposed in
\cite{JulliandTSI01,Masson00} a method that relies on a
partitioning of the transition system into several parts. The
properties are verified on each part separately by an out-of-core
\cite{Toledo99} model checking technique.  As every part is
verified separately from the others, the other parts can be stored
on disks while the part of interest is in the main memory.  We
call \emph{verifiable by parts} the properties that are such that
if they hold on every part of a transition system (whatever the
partitioning), then they hold on the whole transition system. A
sufficient condition $C$ on the B\"{u}chi automaton which accepts
the $\omega$-language of the negation of a property $\neg P$
allows us to decide if $P$ is verifiable by parts. $C$ is a
syntactic condition on the B\"{u}chi automata. Safety and liveness
properties such as $\alws (p \Rightarrow \mynext q), \ \alws (p
\Rightarrow \event q)$ and $\alws (p \Rightarrow r \until q)$ are
verifiable by parts.

The fact that a property $P$ is verifiable by parts does not
depend on the way the parts are constructed.  But, when verifying
$P$ by parts, the fact that $P$ holds on \emph{every} part does
depend on it.  We have proposed a partitioning  based on the
refinement process.

Notice that choosing a partitioning method allows more PLTL properties to be
verifiable by parts.  Some PLTL properties that are not verifiable by parts for
all possible partitionings might become verifiable by parts under the
hypothesis of this particular partitioning.

In this paper, we extend our verification method to liveness
properties on transition systems provided with fairness
assumptions.  To verify a property $P$ under the fairness
assumptions $f$, it is necessary to verify that $f \Impl P$ holds
on the transition system. The problem for verifying $f \Impl P$ by
parts is that even if $P$ is verifiable by parts for all
partitionings, the B\"{u}chi automata of $\neg (f \Impl P)$ does
not in general satisfy the condition $C$.  However, with the fair
refinement based partitioning proposed in the paper, it is enough
that a property $P$ is verifiable by part for all partitionings,
i.e. satisfy $C$, for having $f \Impl P$ verifiable by this
particular partitioning.

\subsection{Related Works}
The component verification of a multiprocess system is achieved by
verifying the properties separately on each component.  Then, the
compositionality with respect to parallel composition ensures that
the properties hold on the whole system
(see~\cite{Clarke89,Kurshan93,kupferman98,Alur&99,Kesten&99}).
Generally, this method allows verifying  safety properties because
a component extracted from its environment is an abstraction of
the parallel system.  For example, \cite{Kurshan93} proves that
$C_1 \| C_2$ satisfies an invariant $I_1 \et I_2$ by proving that
$I_2 \Impl I_1$ holds on $C_1$, and that $I_1 \Impl I_2$ holds on
$C_2$. These methods are called \textit{assume guarantee
paradigm}.  Some methods, like in~\cite{Cheung&97}, use also a
component verification of liveness properties.  In our approach, a
part is not a component of a parallel composition.  Our
partitioned method is \textit{additive} whereas the
compositionality is \textit{multiplicative}.

In this paper we present a fair refinement that  can be compared
with the works on fair simulation presented
in~\cite{Henzinger97,Grumberg94,Kupferman96,Dill91}. The fair
refinement differs from the fair simulation in the following
points:

\begin{itemize}
     \item The fair refinement concerns fair transition systems which
     models action systems, whereas the fair simulation is concerned by
     the ``state systems'' modeled as Kripke structures.  In the fair
     refinement, fairness assumptions are expressed on transitions.  In
     Kripke structure, fairness assumptions are expressed on states.

     \item The fair refinement is a fair $\tau$-simulation because we add
     new actions in the refined system which are not observable
     ($\tau$-actions) in the abstract one.

     \item The fair refinement is a state simulation,  as is the fair simulation,
     but it is also an action simulation, i.e. the sequence of abstract
     actions is a $\tau$-simulation of the sequence of refined actions. This means that the
     refined sequence of actions is identical to the sequence of abstract
     actions, in which finite sub-sequence of $\tau$-actions are
     interleaved.

     \item The verification of the fair refinement is linear in the size of
     the refined system, but the verification of the fair simulation is
     polynomial~\cite{Henzinger97}.
\end{itemize}

Our method can be combined with the parallel composition of
components.  We have proved in~\cite{BellegardeZB02} that the
parallel composition is compatible with the refinement, i.e. if
$C_1$ is refined by $C_3$ then $C_1 \parallel C_2$ is refined by
$C_3 \parallel C_2$.

\mbox{}\par \textsl{Paper Organization.}
The paper is organized as follows.  The preliminaries
(Section~\ref{sec:prelim}) presents the notation and the concepts of fair
transition systems, PLTL and B\"{u}chi automata.  After a presentation of
our refinement relation in Section~\ref{sec:refinement}, we explain the
partitioned model checking technique in Section~\ref{sec:mod-mod-check}.
Section~\ref{sec:pmufa} studies the extension of the partitioned
verification principles to fair transition systems.
Section~\ref{sec:example} gives an example of an application to this
technique, and some experimental results are discussed in
Section~\ref{sec:expResults}.  Finally, we situate our concerns and give
some ideas about future works in Section~\ref{sec:conlu}.

\section{Preliminaries}
\label{sec:prelim}

\subsection{Transition systems}
\label{subsec-behav}
We introduce a finite set of \textit{variables} $x\in\ensvar$ with their
respective finite domains $\domaine_x$.  Let $\enspropat_\ensvar \egdef \{x
= v \tq \ x \in \ensvar, \ v \in \domaine_x\}$ be a set of atomic
propositions over $\ensvar$.

\begin{definition}\theoname{Transition System}
\label{def-TS}
Let $\ensact $ be a nonempty alphabet of labels (names of actions).  A
transition system $\TS \egdef \TSnorm$ interpreted over $V$ has a set of
initial states $S_0$ included in a finite set of states $S$, a labelled
transition relation $\fleche \subseteq S \times Act \times S$ that must be
total, and a state labelling function $L \ : \ S \rightarrow 2^{AP_V}$.
\end{definition}

This is a labelled and interpreted transition system.  A labelled
transition relation $\fleche$ is a set of triples $(s,a,s')$
(written as ``$s \flsup{a} s'$'').  It is an interpreted
transition system because each state is decorated with a set of
atomic propositions.  Notice that the set of atomic propositions
which is associated to a state $s$ must be consistent, i.e. if $x=
v \in L(s)$ and $x= v' \in L(s)$ then $v= v'$.

\begin{remark}
\label{rem:skipLoop}
As the transition relation is total, there can be no deadlock in a
transition system.  If a state $s$ has no successor, a transition $s
\flsup{\mathtt{Skip}} s$ (where $\mathtt{Skip}$ does not belong to $Act$)
is added to obtain a transition system.
\end{remark}

\begin{definition}\theoname{Execution}
\label{def-TSexec} An execution of a transition system $\TSnorm$ is an
infinite alternating sequence $\sigma \egdef s_0 \flsup{a_0}s_1 \flsup{a_1}
s_2 \cdots s_i \flsup{a_i}s_{i+1} \cdots$ of states and actions such that
$s_0 \in S_0$ and for every $i\ge 0$, we have $s_i \flsup{a_i} s_{i+1} \in
\fleche$ \cite{Kurshan&02}.
\end{definition}

We denote by ${\Sigma}_{\TS}$ the set of all the executions of a transition
system $\TS$.

\begin{definition}\theoname{Fragment of an execution}
\label{def-portexec}
We say that $\sigma'$ is a fragment of an execution $\sigma$ (written as
$\sigma' \subset \sigma$) if the sequence of transitions (finite or
infinite) executed in $\sigma'$ is also executed in $\sigma$.
\end{definition}

We note $\transe{s}{*}{s'}$ (resp. $\transe{s}{+}{s'}$) the
fragments of executions leading from $s$ to $s'$ by executing zero
(resp. one) or many transitions.

\begin{definition}\theoname{Cycle}
\label{def-cycle}
A cycle of a transition system $\TS$ is a finite fragment $c \egdef s_0
\flsup{a_0} s_1 \flsup{a_1} \cdots \flsup{a_{n-1}} s_n$ of an execution of
$\TS$, such that $s_n= s_0$ and for all $0 \leq i,j < n$, if $i \neq j$
then $s_j \neq s_i$.
\end{definition}

We extend a finite fragment $\sigma' \egdef s_i\flsup{a_i}s_{i+1}\cdots
s_{k-1}\flsup{a_{k-1}}s_k$ to an execution $\sigma \egdef
s_i\flsup{a_i}s_{i+1} \cdots
s_{k-1}\flsup{a_{k-1}}s_k\flsup{\mathtt{Skip}}s_k\flsup{\mathtt{Skip}}s_k
\cdots$ and we call it an \emph{extension}.

\begin{definition}\theoname{Trace of (a fragment of) an execution}
\label{def-trace}
The trace of an execution or a fragment of an execution $\sigma$, written
as $tr(\sigma)$, is the sequence of the labels of the transitions executed
in $\sigma$.
\end{definition}

Let $\TS_{2}$ be a transition system.  The $\tau$-transition system of
$\TS_{2}$ on $Act_1$, written as $\tau$-$\TS_2$, is a transition system
identical to $\TS_2$ where the names of the actions of $Act_2$ which do not
belong to $Act_1$ are named $\tau$.

\begin{definition}\theoname{$\tau$-transition system}
\label{def:tau-TS}
Let $Act_1$ bet a set of actions.  Let $\TS_2= \TSnormind{2}$ be a
transition system, such that $Act_1 \subseteq Act_2$.  We call
$\tau$-transition system of $\TS_2$ on $Act_1$, the transition system
$\tau$-$\TS_2 =\tuple{{S_0}_2,S_2, {Act_1}_{\tau},
{\flsup{}_2}_{\tau}, L_2}$ such that ${Act_1}_{\tau} = Act_1 \cup \{
\tau \}$, and the relation ${\flsup{}_2}_{\tau}$ is defined as:
\begin{itemize}
\item[]${\flsup{}_2}_{\tau} \egdef \{s_2 \flsup{\tau} s_2' \tq
s_2 \flsup{a_2} s_2' \in \flsup{}_2 \ \wedge\  a_2 \in Act_2
\wedge a_2 \not\in Act_1 \} \cup \{s_2 \flsup{a_1}
s_2' \tq s_2 \flsup{a_1} s_2' \in \flsup{}_2 \ \wedge\  a_1 \in
Act_1 \}$.
\end{itemize}
\end{definition}

Any cycle $c$ such that $tr(c) \in \tau^*$ is called a $\tau$-cycle.

\subsection{Fair transition systems }
\label{subsec-FTS}
We note $\mathit{Inf_s}(\chem)$ the set of states occurring infinitely
often in an execution $\chem$.  We note $\mathit{Inf_t}(\chem)$ the set of
transitions occurring infinitely often in $\chem$.  We note $In(T)$ the set
of states which are source of transitions in a set of transitions $T$.
That is,
\begin{itemize}
\item $\mathit{Inf_s}(\chem)\egdef \{\s \; |\; \sigma= s_0 \flsup{a_0}s_1
  \cdots s_i \flsup{a_i}s_{i+1} \cdots \wedge \ s = \s_i\
\textrm{for infinitely many} \ i\}$,

\item $\mathit{Inf_t}(\chem)\egdef \{\ t \; |\; \sigma= s_0 \flsup{a_0}s_1
  \cdots s_i \flsup{a_i}s_{i+1} \cdots \wedge \ t = s_i
\flsup{a_i} s_{i+1} \ \textrm{for infinitely many} \ i\}$,

\item $In(T)\egdef \{\s \; |\; \s \flsup{a} \s' \in T \}$.
\end{itemize}

Notice that in~\cite{Manna&95}, fairness requirements are
expressed on transitions.  Here, we express fairness requirements
on actions.  Fair transition systems model fair action systems.
Therefore, the user expresses fairness constraints in the form of
fair actions.  In a fair transition system, a fair action is
expressed by a set $F_i$ of the  transitions which  have the name
of this action as a label.

\begin{definition}\theoname{Fair transition system}
\label{def-FTS} Let $\TS = \TSnorm$ be a transition system.  Let
$\enstranseq$ be a set of fairness constraints $\{F_1, F_2,
\cdots, F_m \}$ where every $F_i \subseteq \fleche$ is a set of
the transitions  expressing one fairness constraint.  The fair
transition system $\FTS$ is the tuple $\FTSnorm$ (often written as
$\tuple{\TS, \enstranseq}$) such that
\begin{itemize}
     \item[(a)] $(s_1 \flsup{a} s_1' \in F_{i} \ \et \ s_2 \flsup{b} s_2'
     \in F_{i}) \Impl (a=b)$,

     \item[(b)] $(\transe{s}{+}{s'}\in \to^+\ \et \ \transe{s}{a}{s'}\in
     F_i)
     \Impl \exists(s_1,s'_1) \cdot (\transe{s}{+}{s'} =
     s\flsup{*}s_1\flsup{a}s'_1\flsup{*} s' \et \transe{s_1}{a}{s'_1} \in
     F_i)$,

     \item[(c)] $(s \flsup{a} s' \in F_i \ \et \ s \flsup{a} s'' \in
     \ \enstrans) \Impl (s' = s'')$.
\end{itemize}

\end{definition}

Let us comment on the three constraints (a), (b) and (c).
\begin{itemize}
     \item[(a)] Each element $F_i$ of $\enstranseq$ expresses one fairness
     assumption involving one fair action.  Therefore, all the
     transitions of $F_i$ have the same label.

     \item[(b)]     For each transition $s \flsup{a} s'$ of $F_i$, there does not exist a
     fragment of execution beginning in $s$ and ending in $s'$  that
     does not contain a fair transition of $F_i$. This constraint
     allows expressing  fairness as a PLTL property (see
     Section~\ref{subsec:PLTL})

     \item[(c)]  We require determinism for a fair action since it comes from
     the environment.  So, for each transition $s
     \flsup{a} s'$ of $F_i$,  there
    does not exist another transition $s \flsup{a} s''$, such that $s' \neq
    s''$. This constraint allows verifying the
     refinement in linear time (see
     Section~\ref{subsec:complexVerifRaft})
\end{itemize}

We call \emph{fair transition} a transition of a set $F_i \in
\enstranseq$.

\begin{definition}\theoname{Fair execution (computation)}
\label{def-FTSpath} An execution of $FTS = \langle TS,
\enstranseq\rangle$ (also called a \emph{computation} in
~\cite{Manna&95}), obeys the fairness requirements $\enstranseq$,
which means that: $(i \in [1..m] \ \et \ s \in Inf_s(\sigma) \ \et
\ s \in In(F_i)) \Rightarrow \exists (s', \ a, \ s'')\cdot(s'
\flsup{a} s'' \ \in F_i \ \et \ s' \flsup{a} s'' \in
Inf_t(\sigma))$.
\end{definition}

The set ${\Sigma}_{FTS}$ only contains computations.  In other
words, an execution $\chem$ is fair if it is true that ``if some
transition of any $F_i$ is enabled infinitely often by $\sigma$,
then some transition of $F_i$ is taken infinitely often by the
execution $\sigma$''.

The executions of a transition system which do not satisfy
fairness requirements are called \emph{unfair executions}.

\begin{property}
\label{lem-con-fts1}
An execution $\sigma$ that contains infinitely often two states $s$ and
$s'$ which are in relation by a fair transition $\transe{s}{a}{s'}$ in
$F_i$ is a computation.  So, $(s \in Inf_s(\sigma) \et s' \in Inf_s(\sigma)
\et \transe{s}{a}{s'} \in F_i) \Impl \exists (s_1,s'_1) \cdot
(\transe{s_1}{a}{s'_1} \in F_i \et \transe{s_1}{a}{s'_1} \in
Inf_t(\sigma))$.
\end{property}

\begin{proof}
Let $\sigma$ be an execution that contains $s$ and $s'$ infinitely
often. By the constraint~(b) in Definition~\ref{def-FTS}, $\sigma$
contains a fair transition labelled by $a$ infinitely often.
Therefore, $\sigma$ is a computation.
\end{proof}

\begin{definition}\theoname{Fair Exiting Cycle}
\label{def-faircycle} Let $c= s_0 \flsup{a_0} s_1 \flsup{a_1}
\cdots \flsup{a_{n-1}} s_0$ be a cycle of a fair transition
system.  The cycle $c$ is a fair exiting cycle if there exists a
transition $t = s_i \flsup{a} s'$ of a fairness constraint $F_j$,
such that $s' \neq s_i$, with $i \in [0 \cdots n]$ and $j \in [1
\cdots m]$.
\end{definition}

We call such a transition $t$ an exit transition for a cycle $c$.
By this definition we deduce that the computations of a system do
not run around fair exiting cycles infinitely many times because
they must take exit transitions.

\subsection{Gluing Invariant}
\label{subsec-Inv}
Let $SP_V$ be a set of state propositions over $V$ defined by the grammar
\begin{itemize}
\item[] $p ::= ap\; |\; p \ou p \; |\; \neg p$ where $ap \in AP_V$.
\end{itemize}

\begin{definition}\theoname{Validity of a state proposition}
\label{def-satisf}
A state proposition $p\in SP_V$ is \textit{valid}\footnote{we also say that
``$p$ holds on $s$''} for a state $s$ of a (fair) transition system
(written as $s\sat p$) if
\begin{itemize}
\item $s\sat ap$ iff $ap\in L(s)$,

\item $s\sat p_1 \vee p_2$ iff $s\sat p_1 $ or $s\sat p_2$,

\item $s\sat \neg p$ iff it is not true that $s \vDash p$, written as $s
\nvDash p$.
\end{itemize}
\end{definition}

Let $V_1$ and $V_2$ be respectively the sets of variables of two transition
systems $\TS_1$ and $\TS_2$.  Let $SP_{V_{12}}$ be a set of state
propositions over $V_1$ and $V_2$ defined by the following grammar:
\begin{itemize}
\item[] $q ::= ap_1\; |\; ap_2\;|\; x_1 = x_2 \;|\; q \ou q\;|\; \neg q$ where
$ap_1 \in AP_{V_1},\ ap_2 \in AP_{V_2},\ x_1 \in V_1 \ and \ x_2 \in V_2$.
\end{itemize}

\begin{definition}\theoname{Validity of a state proposition on a pair
of states}
\label{def-satisf2}
A proposition $q \in SP_{V_{12}} $ is valid for a pair of states (written
as $(s_1,s_2) \models q$) if
\begin{itemize}
     \item $(s_1,s_2) \models ap_1$ iff $ap_1\in L_1(s_1)$,

     \item $(s_1,s_2) \models ap_2$ iff $ap_2\in L_2(s_2)$,

     \item $(s_1,s_2) \models x_1 =x_2$ iff \\ $\exists v \cdot (v \in D_{x_1} \
     \et \ v \in D_{x_2} \ \et \ (x_1 = v) \in L_1(s_1)\ \et \ (x_2= v) \in
     L_2(s_2) )$,

     \item $(s_1,s_2) \models q_1 \vee q_2$ iff $(s_1,s_2) \models q_1$ or
     $(s_1,s_2) \models q_2$,

     \item $(s_1,s_2) \models \neg q$ iff it is not true that $(s_1,s_2)
     \models q$.
\end{itemize}
\end{definition}

We call gluing invariant, a state proposition of $SP_{V_{12}}$ which
expresses how the variables from the abstract and the refined transition
systems are linked together.

\begin{definition}\theoname{Gluing invariant}
\label{def-gluinv} A state proposition $\invcol \in SP_{V_{12}}$
is a \emph{gluing} invariant of two (fair) transition systems
$(F)\TS_1$ over $V_1$ and $(F)\TS_2$ over $V_2$, if $\invcol$ is
an invariant on $S_1\times S_2$ (i.e. $(s_1,s_2) \sat \invcol$ for
all pairs $(s_1,s_2)$ of $S_1\times S_2$), and $\invcol$ is a
total function from $S_2$ to $S_1$.
\end{definition}

We require  $\invcol$ to be a total function from $S_2$ to $S_1$
because it allows partitioning $\TS_2$ (see
Section~\ref{subsec:gluingRelAndPartition}).

\subsection{PLTL}
\label{subsec:PLTL} Here, we define all future PLTL formulas with
the two temporal operators, Next ($\mynext$) and Until ($\until$).
We also use the following notations: $\event P$ (eventually $P$)
defined as $true \ \until \ P$, $\alws P$ (always $P$) defined as
$\neg \event \neg P$, and $P_1 \Impl P_2$ defined as $\neg P_1 \ou
P_2$.  The composition of the temporal operators $\alws \event$
means \textit{infinitely often}.

In order to verify PLTL formulas on a (fair) transition system
$(F)\TS$, we extend Definition~\ref{def-satisf} to the PLTL
semantics on the executions (or \textit{computations}) in a
standard way.

\begin{definition}\theoname{PLTL}
\label{def_pltl}
Given PLTL formulas $P,P_1,P_2$ and an execution $\chem$ (or
\textit{computation}), we define $P$ to be valid at the state $s_j$, $j
\geq 0$, on an execution $\chem= s_0 \flsup{a_0} s_1 \flsup{a_1} \cdots,
s_j \flsup{a_j} \cdots$ (written as $(\chem,j)\sat P$) as follows:
\begin{itemize}
     \item $(\chem,j) \sat \propat \text{ iff } \propat \in L(s_j),$

     \item $(\chem,j)\sat \neg P \text{ iff }$ it is not true that
     $((\sigma,j)\models P)$, written as $(\chem,j)\nsat P$,

     \item $(\chem,j)\sat P_1 \ou P_2 \text{ iff } (\chem,j)\sat P_1 \text{
     or } (\chem,j)\sat P_2,$

     \item $(\chem,j)\sat \mynext P \text{ iff}\ (\chem,j+1)\sat P,$ \item
     $(\chem,j)\sat P_1 \until P_2 $ $\text{ iff } \exists k \cdot ( k \geq
     j \et (\chem,k)\sat P_2 \et \ \forall i\cdot(j \leq i< k \Impl
     (\chem,i)\sat P_1))$.
\end{itemize}
\end{definition}

When $(\chem,0)\sat P$ we say that ``$P$ holds on
$\chem$'' and we note $\chem \sat P$.

Now, we extend  Definition~\ref{def_pltl} to transition systems.

\begin{definition}\theoname{Validity of a PLTL formula on a (fair)
transition system} \label{def:10} A PLTL formula $P$ is
\textit{valid} for a (fair) transition system $(F)\TS$, written as
$(F)\TS \sat P$, if $\forall \chem\cdot(\chem \in
{\Sigma}_{(F)\TS} \Impl \chem \sat P)$.
\end{definition}

The PLTL allows  expressing  many dynamics properties such as
safety, liveness and fairness.  Fairness assumptions  expressed by
the set $\enstranseq$ are  described by the PLTL formula
\begin{equation}
\label{eq-pltl-fairness}
f \egdef \bigwedge ^m _{i=1} \Box (\Box
\Diamond \bigvee_{s \flsup{a}s' \in F_i } (\bigwedge_ {ap \in
L(s)} ap) \Rightarrow \Diamond \bigvee_{s\flsup{a}s' \in F_i }
(\bigwedge_ {ap' \in L(s')} ap' ))
\end{equation}
which means that if a transition $t = s \flsup{a}s'$ of a fairness
constraint $F_i$ is infinitely often enabled, then a transition of
$F_i$ must be taken infinitely often.  This description is correct
because of Property~\ref{lem-con-fts1}.

It is important to notice that the verification of a PLTL formula is the
same on a transition system $\TS$ and on a $\tau$-transition system
$\tau$-$\TS$.

\begin{property}
\label{lemma:equivTauTS-TS}
$\tau$-$\TS \models P $ iff $\TS \models P$.
\end{property}

\begin{proof}
The sequences of states which compose the executions in
$\Sigma_{\tau-\TS}$ and $\Sigma_{\TS}$ are the same by
Definition~\ref{def:tau-TS}.  Since the PLTL satisfaction in
Definition~\ref{def_pltl} is verified only on the sequences of states
of the executions, Property~\ref{lemma:equivTauTS-TS} holds.
\end{proof}

A property of a system expressed as a PLTL formula is referred to as a
\emph{PLTL property}.

\subsection{B\"uchi Automata and Executions Acceptance}
We associate to a PLTL formula $P$ a B\"uchi automaton denoted
$\AB_P$. The automaton $\AB_P$  accepts all the executions on
which $P$ holds.

\begin{definition}\theoname{B\"uchi automaton}
\label{def:buchi-autom} A B\"uchi automaton is a 5-tuple $\AB =
\ABnorm$ where:
\begin{itemize}

     \item $\ensetatAB$ is a finite set of states ($\initAB \in
     \ensetatAB$ is the initial state),

     \item $\enstransAB_{\AB}$ is a finite set of transitions labelled by
     elements of $SP_V$ : $\enstransAB_{\AB} \subseteq \ensetatAB \times
     SP_V \times \ensetatAB$,

     \item $\accept \subseteq \ensetatAB$ is the set of \emph{accepting
     states} of the automaton.
\end{itemize}
\end{definition}

Similarly to the notion of execution of a transition system, an infinite
alternating sequence of states of $\AB$ and state propositions in $SP_V$ is
called a \emph{run} of $\AB$.  We denote by $\enschem{\AB}$ the set of all
the runs of $\AB$.  A run of $\AB$ is \emph{accepting} if at least one of
the accepting states appear infinitely often in the run.

\begin{definition}\theoname{Execution Acceptance}
\label{def:infinite-path-rec}
An execution $\chem = \transe{s_0}{a_0}{s_1}\cdots
\transe{s_i}{a_i}{s_{i+1}} \cdots \in \enschem{\TS}$ is accepted by a run
$\chemAB = q_0 \flsup{p_0}q_1 \cdots q_i \flsup{p_i} q_{i+1} \cdots \in
\enschem{\AB}$ if
\begin{itemize}
     \item[i)] $\chemAB$ is a run of $\AB$ on $\chem$: $\forall i \cdot ((0
     \leq i \et \q_{i}\stackrel{p_{i}}{\to}\q_{i+1}\in \to_B) \Impl s_i
     \models p_i $),

     \item[ii)] the run is accepting: $\mathit{Inf_s}(\chemAB) \cap \accept
     \neq \emptyset$.
\end{itemize}
\end{definition}

\section{Refinement}
\label{sec:refinement} In this section, we express the refinement
semantics as a relation between fair transition systems because we
want to verify PLTL properties under fairness assumptions during
the development by the refinement process.  We improve the
refinement relation between transition systems defined
in~\cite{bellegarde2000} by the fairness preservation clause in
order to obtain the fair refinement relation.

We define the refinement relation as a state and action simulation
which allows us to exploit a partition of the refined transition
system state space into parts.  With such a partition we are able
to deal with the model checking blow-up by verifying PLTL
properties of the refined system in a partitioned way.  Moreover
the fairness constraints of the environment make some abstract
system behaviors fair, and these fairness constraints are
preserved during the refinement steps.  So, we want to verify PLTL
properties under fairness assumptions in a partitioned way at the
refined level.

\subsection{Gluing Relation and State Space Partition}
\label{subsec:gluingRelAndPartition} In this section, we consider
two fair transition systems $FTS_1= \FTSnormind{1}$ over
$\ensvar_1$ and $FTS_2=\FTSnormind{2}$ over $\ensvar_2$ giving the
operational semantics of a system at two levels of refinement.
The relation between the variables $V_1$ and $V_2$ is defined
using a gluing invariant $\invcol$.  Our goal is to verify that
$FTS_1$ is {\em refined} by $FTS_2$ (written as $FTS_1
\sqsubseteq_f FTS_2$) according to $\invcol$.

In our approach, the refinement main features are as follows.
\begin{enumerate}
     \item The refinement introduces \emph{new} actions, so $\ensact_1
     \subseteq \ensact_2$.

     \item The refinement renames variables, so $\ensvar_1\cap \ensvar_2 =
     \emptyset$.

     \item The refinement introduces a \emph{gluing} invariant,
     $\invcol$.
\end{enumerate}

Before defining the refinement relation, we define the gluing
relation $\mu$ as a total function from $S_2$ to $S_1$. The
refinement relation is the gluing relation restricted by
additional clauses.

\subsubsection{Gluing Relation and State Space Partition}
\label{subsubsec-partition} We define a binary relation
$\mu\subseteq S_2 \times S_1$ allowing us to express the relation
between the states of two transition systems $TS_1$ and $TS_2$.

\begin{definition}\theoname{Glued states}
\label{def-gluing}
Let $\invcol$ be the gluing invariant.  The state $s_2 \in S_2$ is
\emph{glued} to $s_1 \in S_1$ by $\invcol$, written $s_2 \;\mu\; s_1$, if
$(s_1,s_2) \models \invcol$.
\end{definition}

Since $\mu$ is a total function from $S_2$ to $S_1$, we can define an
equivalence relation $\sim_{\mu}$ between states of the refined transition
system.

\begin{definition}\theoname{Equivalence class}
\label{def:eqclass}
Consider a state $s_1 \in S_1$.  $\eqclass{s_1}$ is an equivalence class of
$S_2/_{\sim_{\mu}}$ if, for every state $s_2\in \eqclass{s_1}$, we have
$s_2\;\mu\; s_1$.
\end{definition}

\subsection{Refinement Relation}
\label{subsec-inclusion} In this section, we define the refinement
of two fair transition systems as a particular kind of simulation
and we view it as computations containment.  For that, we restrict
$\mu$ into a function $\rho_f$ which relates a refined fair
transition system to one of its abstractions.

\begin{figure}
\begin{center}
\includegraphics {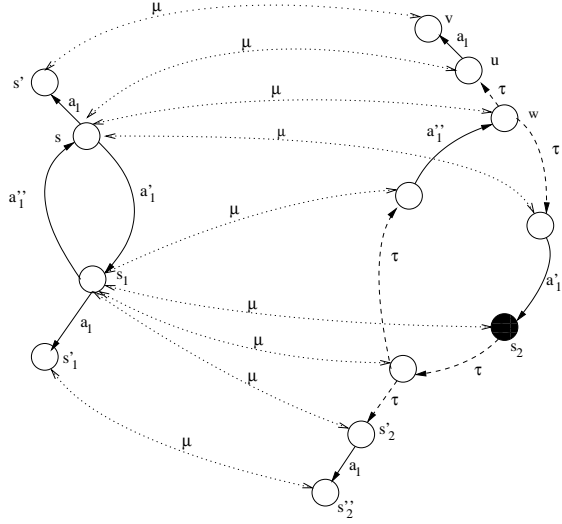}
\end{center}
\caption[Refinement of a fair exiting cycle]{Refinement of a fair
exiting cycle.  The fair transitions at the abstract level are: $t
= s \flsup{a_1} s'$ and $t_1= s_1 \flsup{a_1} s_1'$, such that
$F_{1i} =\{ t, t_1 \}$, $ i \in [1 \ldots m_1]$ and $\enstranseq_1
=\{ F_{1i} \}$.  The transitions $t$ and $t_1$ are the exit
transitions of the cycle at the abstract level.  At the refined
level the state $s_2$ belongs to the set ${S_c}_2$.  So, it is
linked by the relation $\rho_f$ to a state $s_1$ which is a source
state of the fair transition $t_1$ ($t_1 \in F_{1i}$). Therefore
$T_1(s_2) = \{ t_1  \}$. } \label{sc2}
\end{figure}

This relation allows us to distinguish some elements of the state
space partition that we call parts.  This partition is used either
to prove an invariant of a part or to verify state propositions of
a PLTL formula which are verified on a part (see
Section~\ref{sec:mod-mod-check}).  In order to describe the
refinement, we keep the transitions of $FTS_2$ labelled over
$Act_1$ but the \emph{new} ones (from $Act_2\setminus Act_1$)
introduced by the refinement are considered as \emph{non
observable} $\tau$ moves.  These $\tau$ moves hide the transitions
of the parts viewed as transition systems (see Fig.~\ref{sc2}).
Let ${Act_1}_{\tau} \egdef Act_1\union\{\tau\}$.  In the above
parts, it is certainly not desirable that $\tau$ moves take
control forever.  Therefore,   infinite $\tau$-executions are
forbidden.

Let ${S_c}_2$ be the set of the states of $S_2$ (see Fig.~\ref{sc2})
which are targets of a transition labelled by an abstract action, and
are glued with the states of $S_1$ sources of at least a transition in
a fairness constraint $F_i$.  For example, ${S_c}_2$ is the set $\{
s_2, w \}$ in Fig.~\ref{sc2}.  Any cycle at the refined level refining
an abstract fair exiting cycle contains a state of the set ${S_c}_2$.
Let $\sigma$ be a computation which runs around such a cycle $c$
infinitely many often --~this means it reaches a state of ${S_c}_2$.
The computation $\sigma$ must leave $c$ infinitely often in order to
preserve the fairness constraints of the abstract level.
\begin{itemize}
\item[] ${S_c}_2 = \{s_2 \tq \exists (s,a_1').(s \flsup{a_1'} s_2 \in
\flsup{}_2) \ \et \ \exists (s_1, \ s_1 ', \ a_1).(s_1 \flsup{a_1}
s_1 ' \in \bigcup_{i = 1}^{m_1 } F_{1i }) \ \et \ F_{1i} \in
\enstranseq_1 \ \et \ s_2 \mu s_1 \}$.
\end{itemize}

Let $T_1: {S_c}_2 \flsup{ } 2^{{\flsup{}}_1 } $ be a total function.
We denote by $T_{1}(s_{2})$ the set of the abstract fair transitions,
such that their source states are glued with the state $s_2$ by the
relation $\mu$.  For example, $T_1(s_2)$ is the set $\{ t_1 \}$ and
$T_1(w)$ is the set $\{ t \}$ in Fig.~\ref{sc2}.  The computations in
which the state $s_2$ appears infinitely often, execute infinitely
often the transitions of $\flsup{}_2$ which refine the transitions of
$T_1(s_2)$.  \[T_1(s_2) = \{t_1 \tq t_1 = s_1 \flsup{a_1} s'_1 \et t_1
\in \bigcup_{i = 1}^{m_1} F_{1i} \ \et \ F_{1i} \in \enstranseq_1 \
\et s_2\:\mu\:s_1 \}.\]

Let $\tau$-$FTS_2 = \FTStaunormind{2}$ be the fair transition
system resulting from $FTS_2$ and $Act_1$.  Let $s_1$ be a state
of $FTS_1$ $(s_1 \in S_1)$ and $s_2$ be a state of $FTS_2$.  Let
$a$ be a name of action of $FTS_1$, $a \in Act_1$.

\begin{definition}\theoname{$\rho_f$ relation}
\label{def:refinementRel} Let $FTS_1$ and $\tau$-$FTS_2$ be
respectively two fair transition systems provided with a gluing
invariant $\invcol$.  The relation $\rho_f \subseteq S_2 \times
S_1$ is defined from $\invcol$ as the greatest binary relation
included into $\mu$  satisfying the following clauses:
\begin{enumerate}
\item \textbf{strict refinement of abstract transition}
\label{clause-raf1}

$(s_2 \:\rho_f \: s_1 \wedge s_2 \flsup{a}_{2\tau} s'_2 ) \Rightarrow \exists
s_1'\cdot(s_1 \flsup{a}_1 s'_1 \ \wedge \ s'_2 \:\rho_f \: s'_1)$,
\item \textbf{$\tau$-divergence freeness} \label{clause-raf2}

$\forall \sigma\cdot(\sigma \in \enschem{\tau -FTS_2} \Impl
\forall m\cdot (m \ \in\ {Act_1}_{\tau}^*\ \Rightarrow \ tr(\sigma)
\not = m\cdot\tau^{\omega}))$,

\item \textbf{stuttering of $\tau$-transitions }
\label{clause-raf3}

$ (s_2 \:\rho_f \: s_1 \:\et\: s_2 \flsup{\tau}_{2 \tau} s'_2 )
\Rightarrow s'_2
\:\rho_f\: s_1 $,

\item \textbf{abstract fairness preservation} (see Fig.~\ref{sc2})
\label{clause-raf4}

$ s_2 \ \rho_f \ s_1 \  \wedge \  s_2 \in {S_c}_2 \ \et \ s_1
\flsup{a_1} s_1' \  \in T_1(s_2)  \ \et \  i \in [1 \ldots m_1] \
\et \ s_1 \flsup{a_1} s_1' \in F_{1i} \ \et \ s_2 \in
Inf_s(\sigma_2) \ \et \ \sigma_2 \in {\Sigma}_{FTS_2} \
\Rightarrow \ \exists (s, s', u, \ v).( s \flsup{a_1} s' \in
F_{1i} \ \et \  u \flsup{a_1} v \in Inf_t(\sigma_2) \ \et \ u
\ \rho_f\ s \ \et \ v \ \rho_f\ s')  $,

\item \textbf{non reduction of the abstract fairness constraints}
(see Fig.~\ref{sc2}) \label{clause-raf5}  $ s_2 \ \rho_f \ s_1 \
\wedge \ s_2 \in {S_c}_2 \ \et \ s_1 \flsup{a_1} s_1' \ \in
T_1(s_2) \Rightarrow \exists (\sigma_2, s_2', s_2'').(\sigma_2 \in
\enschem{FTS_2} \ \et \ s_2' \flsup{a_1} s_2'' \in \flsup{}_2 \
\et \ s_2' \rho_f s_1 \ \et \ s_2'' \rho_f s_1' \ \et \ (s_2 \in
Inf_s(\sigma_2) \Rightarrow s_2' \flsup{a_1} s_2'' \in
Inf_t(\sigma_2)) )$.
\end{enumerate}

\end{definition}

Clauses~\ref{clause-raf1}-\ref{clause-raf3} are already defined in
\cite{bellegarde2000} in order to define the refinement relation
between two transition systems.  In this paper, we have added
Clause~\ref{clause-raf4} in order to define the refinement
relation between fair transition systems.  Notice that the
presence of Clause~\ref{clause-raf2} guarantees the monotonicity
of an iterative construction of the relation $\rho_f$ and, this
way, the existence of this relation.

Clauses~\ref{clause-raf1}-\ref{clause-raf3} of
Definition~\ref{def:refinementRel} imply that all the
$\tau$-fragments (i.e. sequences of $\tau$-transitions) are finite
and are followed by a transition labelled in $Act_1$.  Therefore,
$\tau$-livelocks are forbidden.  In fair transition systems, the
executions which run around $\tau$-cycles infinitely many times
without taking their exit transition infinitely often, are called
$\tau$-executions. New fairness assumptions must be introduced in
the refined level in order to forbid $\tau$-executions.

The aim of Clause~\ref{clause-raf4} and Clause~\ref{clause-raf5}
is to preserve the abstract fairness constraints.  At the abstract
level, fairness constraints express restrictions in some
executions of the system. The excluded executions are those which
run around fair exiting cycles infinitely many times without
taking their exit transitions infinitely often, they are not
computations.

Clause~\ref{clause-raf4} and Clause~\ref{clause-raf5} are
illustrated by Fig.~\ref{sc2} which represents the refinement of a
fair exiting cycle. In this case, the relation $\rho_f$ between
$\tau$-$FTS_2$ and $FTS_1$ satisfies Clause~\ref{clause-raf4} if
and only if every execution $\sigma$ of $\tau$-$FTS_2$ in which
the state $s_{2}$ (or the state $w$)occurs infinitely often
($s_{2}$ belongs to a fair exiting cycle) activates infinitely
often a transition of $\tau$-$FTS_2$ which is simulated
\footnote{we say that a transition $s_2 \flsup{a} s_2'$ is
simulated by a transition $s_1 \flsup{a} s_1'$ when $s_2 \rho_f
s_1$ and $s_2' \rho_f s_1'$.} by a transition in $F_{1i}$ (because
$t_1 \egdef \transe{s_1}{a_1}{s'_1} \in F_{1i}$). Therefore,
computations must activate infinitely often either the transition
$s_2' \flsup{a_1}s_2''$ (see Fig.~\ref{sc2}) (this transition is
simulated by the transition $t_1$), or the transition $u
\flsup{a_1} v$ (this transition is simulated by the transition $t$
and $t \egdef \transe{s}{a_1}{s'} \in F_{1i}$).

The relation $\rho_f$ between $\tau$-$FTS_2$ and $FTS_1$ satisfies
Clause~\ref{clause-raf5} if and only if there exists  executions
of $\tau$-$FTS_2$ which reaches infinitely often  the states of
${S_C}_2$, i.e. $s_2$ and $w$ respectively
 (they are simulated by the  states sources of the fair transitions
$t_1$ and $t$ respectively), and activates infinitely often the
transitions of $\tau$-$FTS_2$ which are simulated by $t_1$ and $t$
respectively.

So, Clauses~\ref{clause-raf4} and~\ref{clause-raf5} show that any cycle
$c_2$ of $\tau$-$FTS_2$ which is simulated by a fair exiting cycle $c_1$,
must satisfy the following conditions:
\begin{itemize}
     \item $c_2$ must be a fair exiting cycle,

     \item if the exit transitions of $c_1$ are the transitions $t$ and
     $t_1$ (see Fig.~\ref{sc2}), then the computations which run around $c_2$
     infinitely often, must activate
     infinitely often the transition $u \flsup{a_1} v$ or the transition
     $s_2' \flsup{a_1} s_2''$ because they are simulated respectively by
     $t$ and $t_1$,

     \item the cycle $c_2$ must have at least the same number of exit
     transitions as $c_1$.
\end{itemize}

Clause~\ref{clause-raf4} and Clause~\ref{clause-raf5} ensure that
we have a refinement relation which preserves the abstract
fairness constraints, and which also preserves the PLTL
properties.  It means that, a PLTL property satisfied at the
abstract level is also verified at the refined level.  For this,
it is necessary that each computation of the refined level is
simulated by a computation of the abstract system.  This is
guaranteed by Clauses~\ref{clause-raf4} and~\ref{clause-raf5}.
Abstract fairness assumptions are reformulated and introduced at
the refined level in order to satisfy Clauses~\ref{clause-raf4}
and~\ref{clause-raf5}.

When the fair refinement relation holds between two fair transition systems
$FTS_1$ and $FTS_2$, the executions of $TS_2$ which satisfy fairness
assumptions also satisfy Clause~\ref{clause-raf2} and
Clause~\ref{clause-raf4}.

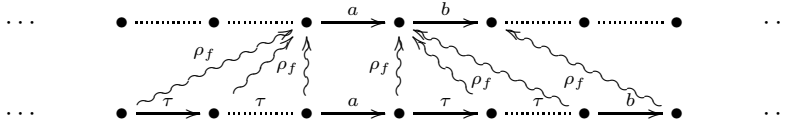
\begin{figure}
\xymatrix{
\cdots & \bullet \ar@{.}[r] & \bullet \ar@{.}[r] & \bullet
\ar[r]^a & \bullet \ar[r]^b &
\bullet \ar@{.}[r] & \bullet \ar@{.}[r] & \bullet & \cdots & \\
\cdots & \bullet \ar[r]^{\tau} \ar@{~>}[urr]^{\rho_f} & \bullet
\ar@{.}[r]^{\tau} \ar@{~>}[ur] & \bullet \ar@{~>}[u]^{\rho_f}
\ar[r]^a & \bullet \ar@{~>}[u]^{\rho_f} \ar[r]^{\tau} & \bullet
\ar@{~>}[ul] \ar@{.}[r]^{\tau} & \bullet \ar@{~>}[ull]^{\rho_f}
\ar[r]^b & \bullet \ar@{~>}[ull]^{\rho_f} & \cdots &   }
\caption{Included computations} \label{pathfig} 
\end{figure}

\begin{remark}
\label{rem:inclComput} Intuitively, the relation $\rho_f$ implies
\textit{computations} containment.  Given $\enschem{FTS_1}$ and
$\enschem{\tau-FTS_2}$, Clauses~\ref{clause-raf1},
\ref{clause-raf2} and \ref{clause-raf3} of
Definition~\ref{def:refinementRel} mean that every execution of
${\Sigma}_{\tau-FTS_2}$ is linked to some execution in
${\Sigma}_{FTS_1}$ (see Fig.~\ref{pathfig}).  In other words, to
every transition labelled $a$ in $\FTS_1$ corresponds a fragment
$\sigma'$ of an execution of $\FTS_2$ where $\sigma'$ is composed
of a sequence of $\tau$-transitions followed by a transition
labelled by $a$.
\end{remark}

\begin{definition}\theoname{Refinement}
\label{def_ref}
A fair transition system $FTS_1= \FTSnormind{1}$ is refined by a fair
transition system $FTS_2= \FTSnormind{2}$ provided with a gluing invariant
$\invcol$, written $FTS_1 \sqsubseteq_f FTS_2$, if $\forall s_2\cdot(s_2
\in S_{0_2}\Rightarrow \exists s_1\cdot(s_1 \in S_{0_1} \et s_2\:\rho_f \:
s_1))$.
\end{definition}

Often initial states are designed ($S_{0_i} \subseteq S_i$).  Then
it is enough that $\rho_f$ holds on the reachable state spaces of
both systems. If there exist $\tau$-cycles, they must be forbidden
by the new fairness assumptions.

\subsection{Analysis of the refinement verification}
\label{subsec:complexVerifRaft} The algorithmic verification of
refinement of finite transition systems can be effectively done by
a joined exploration of the reachable state spaces. We have given
a verification algorithm of the fair refinement in
\cite{chouali&03,ChoualiThese03} (the proof of the algorithm was
also given).

Let $\size{\TS_1}$ and $\size{\TS_2}$ be respectively the sizes of the
transition system and one of its refinement.  Suppose $\mu$ is a
function.  Verifying Clauses~(1),(3) requires a parallel exploration
of the systems $TS_1$ and $TS_2$.  Same goes for Clauses~(4),(5)
because of the constraint (c) in Definition~\ref{def-FTS}.  Therefore,
the complexity is $\mathcal{O}(\size{\TS_1} + \size{\TS_2})$ $\simeq$
$\mathcal{O}( \size{\TS_2})$ (because generally $\size{\TS_2} >
\size{\TS_1}$ ).  For a finite $\TS_2$, verifying Clause~(2) requires
a search for $\tau$-cycles by exploring paths, of $\TS_2$, and by
following fair transitions.  This verification does not change the
complexity ($\mathcal{O}( \size{\TS_2})$).

However, if $\mu$ is not a function, the verification necessitates
a joint enumeration of both systems. The complexity becomes in
$\mathcal{O}(\size{\TS_2} \times \size{\TS_1})$.

In the next section, we give an overview of how to verify PLTL
properties without considering environment fairness constraints,
by a partitioned model checking, as it is presented
in~\cite{Masson00,JulliandTSI01,MassonThese01}.

\section{Partitioned Model Checking}
\label{sec:mod-mod-check} In this Section, we present the main
results of an out-of-core model checking technique that we have
developed in order to face the state explosion problem for the
PLTL model checking.  This technique has been presented
in~\cite{Masson00,JulliandTSI01} for transition systems without
fairness assumptions.

In order to perform model checking on large transition systems, the
partitioned verification technique relies on a simple idea: why not split
the transition system into several smaller pieces, and perform the
verification on each piece separately?  The pieces are called \emph{parts}.
Parts are transition systems as well.  The initial transition system is
called the \emph{global} transition system.

In order to have every transition in one part, the parts are constructed by
partitioning the transitions of the global transition system.  Some states
may belong to two distinct parts: they can be the target state of a
transition $t$ in one part, and the initial state of a transition $t'$ in
another part.

To perform a partitioned verification is to verify a property on each part
separately, and to conclude that it is globally true when it is true on
every part.

Section~\ref{subsec:vbp-props} defines what is a property verifiable by
parts, and exhibits a class of such PLTL properties.
Section~\ref{subsec:part-by-ref} proposes a partitioning of a transition
system based on the refinement.

\subsection{Properties Verifiable by Parts}
\label{subsec:vbp-props}

\subsubsection{Definition of a Property Verifiable by Parts}
\label{subsubsec:def-mod-ver-prop} Consider a transition system
split into a set of parts (transition systems) according to a
partition of its set of transitions. Actually, it is enough that
the parts are obtained by an overlapping of the transitions. Some
PLTL properties have the property to be globally true when they
are true on every part. We call such properties \emph{verifiable
by parts}.

\begin{definition}\theoname{Property verifiable by parts}
\label{def:mod-ver-prop}
Let $P$ be a PLTL property.  Let $\TS$ be a transition system, and let
$\ensmodule$ be a partitioning of $\TS$.  The property $P$ is verifiable by
parts on $\TS$ if
\begin{equation}
\label{eq:def-mod-ver-prop}
 \forall\module \cdot (\module\in
\ensmodule \Impl \module\sat P) \Impl {\TS} \sat P.
\end{equation}
\end{definition}

\begin{remark}
We simply say \emph{$P$ is verifiable by parts}, instead of \emph{$P$ is
verifiable by parts on $\TS$}.
\end{remark}

\subsubsection{A Class of PLTL Properties Verifiable by parts}
\label{subsubsec:PLTL-class}
Before we perform a partitioned verification, we have to make sure that the
properties that we want to verify are verifiable by parts.  That is, we
have to prove~(\ref{eq:def-mod-ver-prop}) on every property.  Let $P$ be a
PLTL property.  Notice that to prove ``if $P$ is true on every part, then
it is true on the global transition system'' is equivalent to prove ``if $P$
is false on the global transition system, then it is false on one part at
least''.  That is, to prove~(\ref{eq:def-mod-ver-prop}) is equivalent to
prove~(\ref{eq:contrap}):
\begin{equation}
\label{eq:contrap}
\neg({\TS}\sat P) \Impl \exists\module \cdot \exists\chem \cdot
(\module\in\ensmodule \et \chem\in\enschem{M} \et \chem\sat\neg P).
\end{equation}

By using B\"{u}chi automata, we give a sufficient condition for
when a PLTL property is verifiable by parts.  Consider a property
$P$, and an execution $\chem$ on which $P$ does not hold. Suppose
that there is a state $s$ in $\chem$ such that every fragment of
$\chem$ starting in $s$ violates $P$.  Then the part containing
$s$ violates $P$ since it contains a fragment of $\chem$ starting
in $s$.  The same idea works when a unique transition is the cause
of such a violation of the property.  Because every state (and
every transition) necessarily belongs to one part, we know that
the property does not hold on the part that contains this state
(or transition).

We define a class $\ABmod$ of B\"uchi automata, and we prove that every
PLTL property whose negation defines a language that is recognized by an
automaton in $\ABmod$ is a property verifiable by parts.  An automaton in
$\ABmod$ has every accepting state leading only to an accepting state, and
there is at most one intermediate state between the initial state and any
accepting state.  Moreover, there is a loop labelled by \texttt{True} on
the initial state.

\begin{definition}\theoname{Class $\ABmod$ of B\"uchi automata}
\label{def:Cmod-class}
Let $\AB = \ABnorm$ be a B\"uchi automaton.  We have $\AB \in \ABmod$ if
\begin{enumerate}
     \item  there is a loop labelled \texttt{True}\ on the initial state:
     $\transe{\initAB}{\mathtt{True}}{\initAB} \in \enstransAB_{\AB}$,

     \item for any run $\chemAB = \ q_0 \flsup{p_0}q_1 \flsup{p_1} q_2
     \ldots q_i \flsup{p_i} q_{i+1} \ldots \ \in \enschem{\AB}$
     \begin{equation}
    \label{eq:def-Cmod}
    \exists k\cdot (k>0 \et \forall i \cdot((0\leq i < k \Impl
    \q_{i}=\initAB) \et (i > k
    \Impl \q_{i}\in \accept))),
     \end{equation}

    \item $\forall(q,p,q') \cdot (\transe{q}{p}{q'} \in \enstransAB_{\AB} \et
     q'\in\accept \Impl \exists (p',q'') \cdot (\transe{q'}{p'}{q''} \in
     \enstransAB_{\AB} \et p \Impl p'))$.\label{cl:formClauseLabelCmod}
\end{enumerate}
\end{definition}

\begin{theorem}
\label{BA1modular} All the PLTL properties whose negation defines a
language that is recognized by a B\"uchi automaton in the class $\ABmod$
are properties verifiable by parts.
\end{theorem}

\begin{proof}
Let $P$ be a PLTL property and let $\TS$ be a transition system on which $P$
does not hold.  We have $\neg (\TS \sat P)$, that is:
\[\exists
\chem\cdot(\chem\in\enschem{\TS} \et \chem\sat\neg P).
\]
Let $\AB_{\neg P}\in\ABmod$ be a B\"uchi automaton that recognizes the
language of $\neg P$.  There exists a run $\chemAB = q_0
\flsup{p_0}q_1 \flsup{p_1} q_2 \ldots q_i \flsup{p_i} q_{i+1} \ldots \
\in \enschem{\AB_{\neg P}}$ of $\AB_{\neg P}$ on $\chem$ on
which~(\ref{eq:def-Cmod}) holds.  With an index $k$ as defined in
Formula~(\ref{eq:def-Cmod}), we consider $\s_{k-1}$ and $\s_{k}$,
respectively the $(k-1)$-th and $k$-th states in the execution
$\chem$.  With any partitioning of $\TS$, the transition
$\s_{k-1}\stackrel{a_{k-1}}{\to} \s_{k}$ necessarily belongs to a part
$\module$, and the state $s_{k-1}$ is reachable with transitions of
$\module$ from an initial state $s'_{0}$ of $\module$.  Let
\[
\sigma' = \transe{s'_0}{a'_0}{s'_1} \cdots \transe{s_{k-1}}{a_{k-1}}{s_k}
\cdots
\]
be a fragment of an execution of $\TS$ such that the suffix
$\transe{s_{k-1}}{a_{k-1}}{s_k} \cdots$ is common to $\sigma$ and
$\sigma'$, and such that all the transitions appearing in the
prefix $\transe{s'_0}{a'_0}{s'_1} \cdots
\transe{s_{k-1}}{a_{k-1}}{s_k}$ of $\sigma'$ are transitions of
$\module$.

Consider the run
\[
\pi' = \transe{q_0}{\mathtt{True}}{q_0} \cdots
\transe{q_0}{\mathtt{True}}{\transe{q_{k-1}}{p_{k-1}}{\transe{q_k}{p_k}{q_{k+1}}}}
\cdots
\]
of $\AB_{\neg P}$ where the suffix $\transe{q_{k-1}}{p_{k-1}}{q_k}\cdots$
is common to $\pi$ and $\pi'$.  Such a run exists as $\AB_{\neg P} \in
\ABmod$ and as $q_{k-1}=q_0$ by construction.  Moreover, the run
$\pi'$ is accepting on $\sigma'$ because it ``stays'' on $q_0$ until it
accepts the suffix $\transe{s_{k-1}}{a_{k-1}}{s_k}\cdots$ the same way it
accepted it in $\sigma$.

There are two cases.
\begin{enumerate}
\item $\chem'$ is an execution of $\module$.
Then $\sigma'$ is accepted by $\AB_{\neg P}$ and so $\neg(\module\sat P)$.

\item $\chem'$ is not an execution of $\module$: there is a state $s_c$ ($c
\geq k$) such that the prefix $\transe{s'_0}{a'_0}{s'_1} \cdots
\transe{s_{k-1}}{a_{k-1}}{s_k} \cdots \transe{s_{c-1}}{a_{c-1}}{s_c}$ of
$\chem'$ is a fragment of an execution
\[
\sigma'' = \transe{s'_0}{a'_0}{s'_1} \cdots \transe{s_{k-1}}{a_{k-1}}{s_k}
\cdots
\transe{s_{c-1}}{a_{c-1}}{\transe{s_c}{\mathtt{Skip}}
{\transe{s_c}{\mathtt{Skip}}{s_c}}} \cdots
\]
of $\module$.  The transition $\transe{s_c}{a_c}{s_{c+1}}$ is not in
$\module$.

Consider
\[
\pi''=\transe{q_0}{\mathtt{True}}{q_0}\cdots \transe{q_c}{p_c}
{\transe{q_{c+1}}{p'_0}{\transe{q'_1}{p'_1}{q'_2}}}\cdots
\]
an accepting run of $\AB_{\neg P}$ where the prefix $q_0\cdots
q_{c+1}$ is common to $\pi'$ and $\pi''$.  As $\AB_{\neg P}\in\ABmod$,
then $\forall j\cdot (j\geq 1\Impl q'_{j}\in\accept)$.  Moreover, from
Clause~\ref{cl:formClauseLabelCmod} in
Definition~\ref{def:Cmod-class}, $p_{c} \Impl p'_{0}$ and $\forall j
\cdot (j\geq 1 \Impl (p'_{j-1}\Impl p'_{j}))$.  As a consequence,
$\forall j\cdot (j \geq 0 \Impl (p_c\Impl p'_j))$.  Thus, $\pi''$ is
an accepting run on $\sigma''$, i.e. $\forall i \cdot (i\geq k-1 \et i
\leq c \Impl s_{i}\models p_{i}) \et \forall j \cdot (j\geq 1 \Impl
s_{c}\models p'_{j})$.  As the execution $\sigma''$ is accepted by
$\AB_{\neg P}$, then $\neg(\module\models P)$.
\end{enumerate}
\end{proof}

\begin{figure}
   \centering
   \renewcommand{\subfigcapskip}{10pt}
   \begin{tabular}{cc}

\subfigure[\label{subfig:aut-nnext}$\mathcal{B}_{\neg\alw(p\Impl\nxt
q)}$]{
       \includegraphics[scale=0.55]{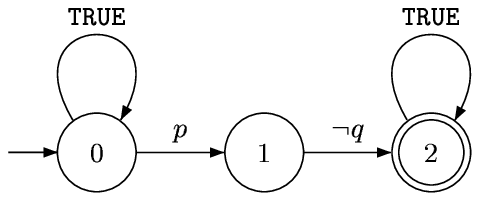}
       }
       &

\subfigure[\label{subfig:aut-nfatalt}$\mathcal{B}_{\neg\alw(p\Impl\evtly
q)}$]{
       \includegraphics[scale=0.55,trim=-20 0 -20 0]{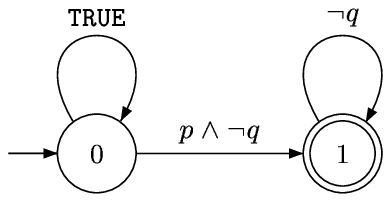}
       }
       \\
       \subfigure[\label{subfig:aut-nuntil}$\mathcal{B}_{\neg\alw(p\Impl
       q\:\until\:r)}$]{
       \includegraphics[scale=0.55]{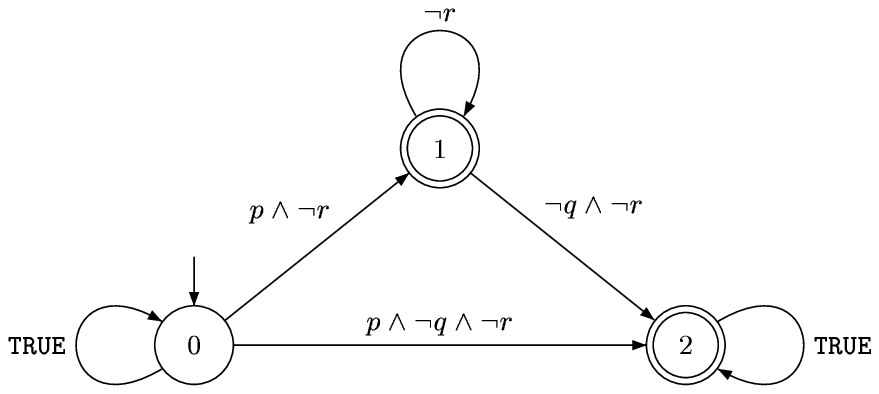}
       }
       &
       \subfigure[\label{subfig:aut-nfairness}$\mathcal{B}_{\neg(\alw(\alw\evtly
       p \Impl \evtly q) \Impl \alw(r \Impl \evtly s))}$]{
       \includegraphics[scale=0.55]{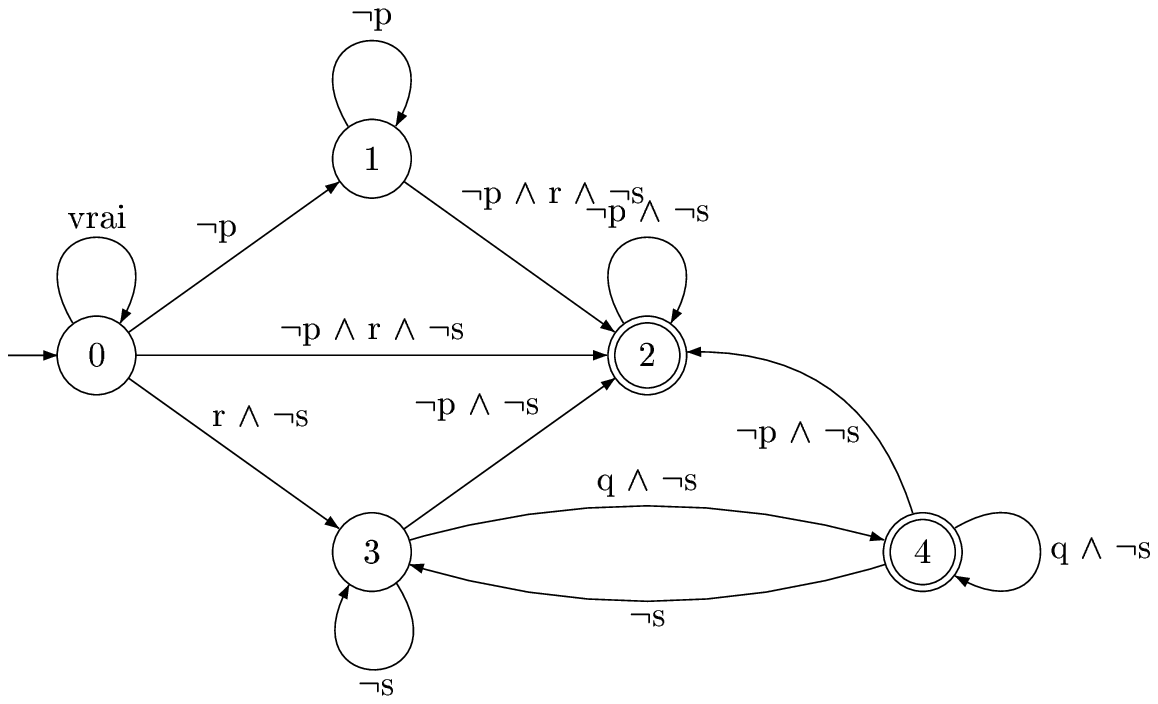}
       }
       \\
  \end{tabular}

      \caption{Some B\"uchi automata}
      \label{fig:vbpprops}
\end{figure}

The class $\ABmod$ contains some safety properties such as $\alw
p$ or $\alw(p\Impl\nxt q)$ (see Fig.~\ref{subfig:aut-nnext}), all
reachability properties such as $\alw\neg p$, and some liveness
properties such as $\alw(p\Impl\evtly q)$ and $\alw(p\Impl
q\:\until\: r)$ (see Fig.~\ref{subfig:aut-nfatalt} and
Fig.~\ref{subfig:aut-nuntil}).  However it does not contain
liveness properties under fairness assumptions such as
$\alw(\alw\evtly p \Impl \evtly q) \Impl \alw(r \Impl \evtly s)$
(see Fig.~\ref{subfig:aut-nfairness}).

\subsection{A partitioning based on refinement}
\label{subsec:part-by-ref} Consider a property $P$ that is
verifiable by parts, and a transition system $\TS$ on which $P$
globally holds.  Actually, it is very likely that with an
inappropriate partitioning $\ensmodule$ of $\TS$, there is a part
$M$ of $\ensmodule$ on which $P$ does not hold.  In this case, we
can not conclude that $P$ is globally true from the partitioned
verification.  In other words,  the fact that a property globally
true is also true on every part depends on the way the global
transition system is partitioned.

As an heuristic to the choice of an appropriate partitioning, we propose to
partition a transition system according to the refinement process.

At every step of the refinement, the specifier introduces new actions and
refines the old ones.  We propose that the specifier also introduces the
PLTL properties that can be observed thanks to these new actions at this
very level of refinement.  The properties verified at the former levels of
refinement need not to be verified again since PLTL properties are
preserved by refinement \cite{ZB2003}.

The new properties are likely to be observed on the
``successions'' of new actions, that is on the $\tau$-executions.
Intuitively, a property that could be observed on a succession of
more than one old action is not a new property.

According to this idea, we propose that the parts contain the
$\tau$-executions of a refined system.  Parts constructed in this
way are called \emph{refinement based parts}.  We have proposed in
\cite{JulliandTSI01} a definition of such parts.  The parts are
built as follows (see Remark~\ref{rem:inclComput}): to every state
of the abstract system corresponds a part in the refined system.
Let $s_1$ be a state of the abstract system.  The part
corresponding to $s_1$ is made of all the transitions of the
refined system that have a state of $\eqclass{s_1}$ as a source
state.  The target state $s'$ of such a transition is
\begin{itemize}
\item either a state itself glued to $s_1$ -- then the transition
represents the occurrence of a new event, or of an old event if it refines
an abstract transition $s_1\flsup{a}s_1$,

\item or a state not glued to $s_1$ -- then the transition represents the
occurrence of an old event, and means that the ``end'' of the module is
reached: $s'$ has no successor in the part other than itself by the virtual
transition labelled \texttt{Skip} (see Remark~\ref{rem:skipLoop}).
\end{itemize}

The definition that we propose in this paper is slightly different
because we are in the context of the refinement of fair transition
systems.  Thus, Definition~\ref{def:module} defines a part of a
refined fair transition system according to the fair refinement
relation $\rho_{f}$. A part is a transition system.

Let us first give the intuition of what are the initial states,
the states and the transitions of such a part.  Consider an
abstract fair transition system $\FTS_1$ and a fair transition
system $\FTS_2$ that refines $\FTS_1$.  Let $s_1$ be a state of
$\FTS_1$.  We define the part $\TS_\module$ corresponding to
$s_1$. We define $Y$ as being the greatest set of states of
$\FTS_2$ that are successors of a state in $\eqclass{s_1}$ by
taking a transition labelled with an abstract action: $Y= \{ \
s'\tq \transe{\s_{2}}{\act}{\s'} \in\ \enstrans_{2}\et\
\s_{2}\in\eqclass{s_1} \et \s'\not\in \eqclass{s_1}\}$. We define
$FS(Y)$ as being the greatest set of states of $\FTS_2$ that are
reachable from a state of $Y$ by taking only fair transitions:
$FS(Y) =\{s_j \tq 1 \leq j \leq n \et \exists \sigma \cdot( \sigma
\in \enschem{\FTS_2} \et (s_0 \flsup{a_0}s_1 \flsup{a_1} \dots
\flsup{a_{i-1}}s_i \dots \flsup{a_{n-1}}s_n) \subset \sigma \et
s_0 \in Y \et \forall i\cdot(s_i \flsup{a_i} s_{i+1} \in
\bigcup_{j = 1}^{m_2 } F_{2j}))\}$.

The initial states of $\TS_\module$ are the states of
$\eqclass{s_1}$ that are either initial states of $\FTS_2$, or
target states of a transition labelled with an abstract action
whose source state is not in $\eqclass{s_1}$.

The transitions of $\TS_\module$ are:
\begin{itemize}
     \item the transitions of $\FTS_2$ that have a state of $\eqclass{s_1}$
     as a source state,

     \item the fair transitions of $\FTS_2$ that have a state of $Y \union
     FS(Y)$ as a source state,

     \item the transitions labelled with \texttt{Skip} that are added as
     loops on the states of $\TS_\module$ which have no successors among the
     states of $\TS_\module$.
\end{itemize}

\begin{definition}\theoname{Refinement Based Part of a Fair Transition System}
\label{def:module} Let $\FTS_{1} = \tuple{S_{0_1}, S_1, \ensact_1,
\enstrans_1, \fdecor_1, \enstranseq_1}$ be a fair transition
system which is refined by a fair transition system $\FTS_{2} =
\tuple{S_{0_2}, S_2, \ensact_2, \enstrans_2, \fdecor_2,
\enstranseq}$ ($\FTS_{1}\sqsubseteq_{f} \FTS_{2}$).  Consider
$s_1\in S_1$ and $\eqclass{s_1}$, an equivalence class of
$S_{2}/_{\sim_{\mu}}$. The part based on $\eqclass{s_1}$ is a
transition system $\TS_{\module} = \TSnormind{\module}$ such that:

\begin{itemize}
     \item $\ensetat_{0_{\module}} = \{\s_{2}\in\eqclass{s_1} \tq \s_{2}\in\
     \ensetat_{0_{2}} \ou \exists\s \cdot \exists\act \cdot
     (\transe{\s}{\act}{\s_{2}} \in \enstrans_{2} \et \ \s \not\in
     \eqclass{s_1})\}$.

     \item $\ensetat_{\module} = \{\s_{2}\in\eqclass{s_1} \}\union Y \union
     FS(Y)$.

     \item $\enstrans_{\module} = \{ \transe{\s_{2}}{\act}{\s'} \in\
     \enstrans_{2}\ \tq \s_{2}\in\eqclass{s_1}\} \ \union \
     \{\transe{\s}{\act}{\s'} \in \flsup{}_2\ \tq s \in (Y \ \union \ FS(Y)) \
     \et \ s' \in FS(Y) \} \ \union \ \{ s \flsup{Skip} s \tq s \in
     \ensetat_{\module} \ \et \ \forall(s',a)\cdot(s \flsup{a}s' \in
     \flsup{}_2\ \Rightarrow s' \not \in \ensetat_{\module}) \}$.

     \item $\ensact_{\module}$ is the restriction of $\ensact_{2}$ to the labels
     of $\enstrans_{\module}$, augmented with \texttt{Skip}.

     \item $\fdecor_{\module}$ is the restriction of $\fdecor_{2}$ on the states
     of $\ensetat_{\module}$.
\end{itemize}
\end{definition}

Notice that there are at most as many parts in the refined transition
system as states in the abstract transition system.  As a consequence, the
number of states of the abstract model can be used as a parameter for the
user to control either the number, or the size of the parts to be
model-checked.

\begin{remark}
The properties that are not verifiable by parts for all possible
partitionings, but only relatively to the  partitioning as
presented in Definition~\ref{def:module}, are called
\emph{verifiable by refinement based parts}.
\end{remark}

   \subsubsection*{Example of refinement based partition according to 
   Definition~\ref{def:module}}

\begin{figure}
\begin{center}
\includegraphics{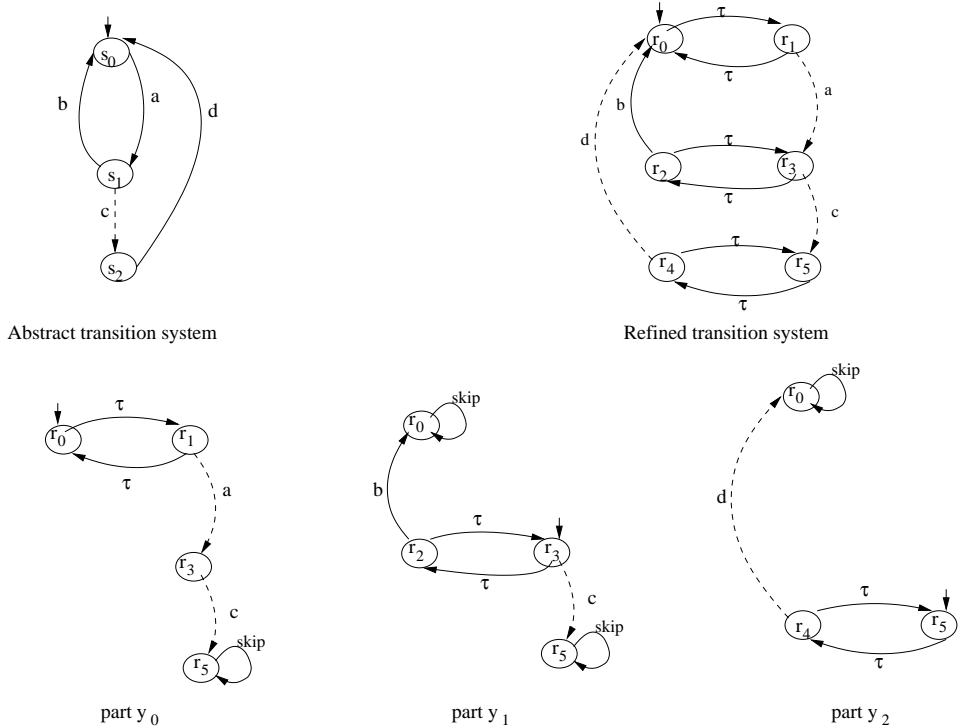}
\end{center}
\caption{Example of a refinement based parts} \label{figexplde46}
\end{figure}

   In Fig.~\ref{figexplde46}, we present an  example of three parts $y_0$, $y_1$, and $y_2$ obtained according to
   Definition~\ref{def:module} from  the refined
   transition system. We suppose that the states of the refined system and the abstract system
   are glued as : $r_0$ and $r_1$ with $s_0$,  $r_2$ and $r_3$
   with   $s_1$, and  $r_4$ and $r_5$ with $s_2$. The fair
   transitions are represented by dashed arrows in
   Fig.~\ref{figexplde46}.

For example the part $y_0$ is the transition system composed of
the states  $EC(s_0) =\{r_0, \ r_1\}$, $Y = \{ r_3 \}$ and $FS(Y)=
\{ r_5 \}$. It has one initial state $r_0$. It contains all
$\tau$-transitions between the states of $EC(s_0)$, the exit
transition $r_1 \flsup{a} r_3$, and the fair transition $r_3
\flsup{c} r_5$. As this transition system deadlocks in $r_5$, we
extend it with the transition $r_5 \flsup{Skip} r_5$.

Next section describes how the partitioned verification applies
for the verification of PLTL properties under fairness
assumptions.

\section{Partitioned Model Checking under Fairness Assumptions}
\label{sec:pmufa} This section contains the main contribution of
the paper.  We study the verification by the partitioned model
checking of PLTL properties when the description of the
environment uses  fairness constraints.  We show that partitioned
model checking on transition systems can be used under fairness
assumptions.

We do not include the fairness of the environment in the
transition system, but it is integrated as an assumption of the
property to verify.

 The verification by model checking of a PLTL
property $P$ under fairness assumptions expressed by a PLTL
formula $f$, on a transition system $TS$, supposes verifying by
model checking the property $f \Rightarrow P$ on $TS$. The
question ``does $TS$ provided with fairness assumptions $f$
satisfies $P$ ?'' is written as: $TS \models f \Rightarrow P$~?

The B\"uchi automaton of $\neg(f \Rightarrow P)$ does not satisfy
the sufficient condition allowing its verification in a
partitioned way (see Section~\ref{sec:mod-mod-check} and Fig.
3d). However, we prove that the verification of the formula $f
\Rightarrow P$ becomes verifiable by refinement based parts (see
Definition~\ref{def:module}) , i.e. $\forall M.(M \textit{ is a
refinement based part } \Rightarrow M \models f \Rightarrow P)
\Rightarrow TS \models f \Rightarrow P$ when $P$ is verifiable by
parts. This is equivalent to say  that if, for all $M$ such that
$M$ is a refinement based part, each computation of $M$ satisfies
$P$ then all the computations of $TS$ satisfies $P$. Indeed, since
the unfair executions  of $M$ and $TS$ satisfy the formula $f
\Rightarrow P$ because they do not satisfy $f$,  we deduce that
the verification of $f \Rightarrow P$ on $M$ (refinement based
part) and on $TS$ requires only the verification of $P$ on the
computations (fair executions) of $M$ and $TS$.

Now, let us prove that each computation of $TS$ is a concatenation
of fragments which are all prefix of computations in  refinement
based parts.

\begin{lemma} \label{principal-lem}
Suppose $FTS_1 \sqsubseteq_f FTS_2$. Each computation $\sigma$ of
$FTS_2$ can be decomposed into fragments which are prefix of
computations in  refinement based parts of $FTS_2$. More
precisely, such a computation is, either a suffix of $\sigma$, or
is a fragment of $\sigma$ which ends by a finite sequence of fair
transitions followed by an infinite sequence of $skip$
transitions.
\end{lemma}

\begin{proof}
As the fair refinement is a $\tau$-simulation of $\tau$-$FTS_2$ by
$FTS_1$, each computation $\sigma$ of $\tau$-$FTS_2$ is such that
\begin{itemize}
\item[] $\sigma = \ s_0 \flsup{\tau *}s_{i_1 - 1}
\flsup{a_{i_1}}s_{i_1}\flsup{\tau *} s_{i_2 -1} \flsup{a_{i_2}}
s_{i_2} \flsup{\tau *} s_{i_3 - 1} \flsup{a_{i_3}} s_{i_3} \ldots$
\end{itemize}
where $a_{i_j}$ are abstract actions.  The decomposition by
Definition~\ref{def:module} is such that each finite fragment of
$\sigma$, $\varphi_f = \ s_{i_{j - 1}} \flsup{\tau *}s_{i_j - 1}
\flsup{a_{i_j}}s_{i_j}$ is a prefix of a computation of a refinement
based part such as, either
\begin{itemize}
\item $\sigma' =  \ s_{i_{j - 1}} \flsup{\tau *}s_{i_j - 1}
\flsup{a_{i_j}} s_{i_j} \flsup{c*}s_c \flsup{\tau *}s_{i_j - 1}
\flsup{a_{i_j}} s_{i_{j}} \flsup{c*}s_c \ldots$, where $c \in
Act_1 \cup \{ \tau \}$ and $s_{i_j} \flsup{c*} s_c$ is a finite
sequence of fair transitions and states $s_c$ are not source
 states of a fair transition -here $\sigma'$ refines
a suffix of a computation of a global system in the abstract level
which run around a cycle infinitely many times -, or

\item $\sigma'' =  \ s_{i_{j - 1}} \flsup{\tau *}s_{i_j - 1}
\flsup{a_{i_j}}s_{i_j} \flsup{c*}s_c \flsup{Skip}s_c
\flsup{Skip}s_c \ldots$.
\end{itemize}

Obviously  $\sigma'$ is a computation since $\sigma$ is a
computation and $\sigma'$ is a suffix of $\sigma$. Also
 $\sigma''$ is a computation since, by construction, it is a fragment of
the computation $\sigma$ prolongated by a finite sequence of fair
transitions followed by an infinite sequence of $skip$
transitions.
\end{proof}

\subsection*{Correctness of the Partitioned Model Checking under Fairness Assumptions}
In this section, we show that model checking by parts under
fairness assumptions is sound. For that it is  necessary  that the
parts are obtained according to Definition~\ref{def:module}.

\begin{theorem}
\label{th-verifpart}

Let $FTS = \tuple{TS, \enstranseq}$ be a fair transition system which
refines an abstract fair transition system.  Let $\ensmodule$ be the
set of refinement based parts accordingly to
Definition~\ref{def:module}.  Let $f$ be the PLTL formula which
expresses the fairness assumptions of $FTS$.  If $P$ is a PLTL formula
such that $\AB_{\neg P} \in \AB_{mod}$ (as such, it is verifiable by
parts on $TS$), then the property $f \Rightarrow P $ is verifiable by
refinement based parts on $\TS$.

\end{theorem}

\begin{proof}
Recall that in order to prove that the formula $f \Rightarrow P$ is
verifiable by refinement based parts on $\TS$, we must prove the
following: if, for all $M$ such that $M$ is refinement based part,
each computation of $M$ satisfies $P$ then all the computations of
$TS$ satisfies $P$.

Let $\sigma$ be a computation of $\TS$.  By Lemma~\ref{principal-lem},
each fragment of a computation $\sigma$ prefixes a computation like
$\sigma'$ or $\sigma''$ of a refinement based part.  Therefore all the
computations of a refinement based parts are extensions of all the
fragments of the computations belonging to a global system.

 In  the proof of Theorem 4.4 in Section 4.1.2  it is shown  that  when a
property $P$ belong to the class of properties verifiable by
parts, and if  the extensions of all the fragments of an execution
$\sigma_i$ (in a global system) satisfy $P$, then $P$ holds on
$\sigma_i$.

So, when each computation of a refinement based part satisfies $P$
and since $P$ is verifiable by parts, we conclude by Theorem 4.4
that $P$ holds on $\sigma$. Which means that $f \Rightarrow P$ is
verifiable by refinement based parts on $\TS$.
\end{proof}

\section{Example of the Protocol T=1}
\label{sec:example} In this section, we present the example of the
protocol T=1 \cite{teg1} in order to illustrate how to verify PLTL
properties under fairness assumptions in a partitioned way. We
also defined how to express the fairness of an environment in a
$B$ event system.  We give the $B$ event systems enriched with
fairness assumptions, of the protocol at the abstract and at the
refined level. We also give the fair transition system which is
the semantics of the abstract event system and the fair transition
system which expresses the semantics at the refined level.

We also use this example to show that without using fair
refinement to obtain the set of parts, the PLTL properties under
fairness assumptions are not verifiable by parts.

\subsection{Abstract Specification under Fairness Assumptions}

Figure~\ref{facfigbspec1} represents the abstract $B$ event system
of a half duplex communication protocol between a chip integrated
card and a card reader. At this level of specification, we
consider only the alternation of exchange of messages between the
chip card and the reader.

Figure~\ref{fig1} represents the abstract transition system of the
protocol.  In this figure, each state is decorated with the value
of the state variables. $Cstatus_1$ indicates if the chip card is
inserted or not in the reader. $Sender_1$ indicates the device
which will send the next message, the chip card or the reader. The
character '?'  or '!'  in the reader indicates respectively that
the reader is the receiver or the sender device.  The state
labelling function, called $L_1$ is the following: $L_1 =\{$
\begin{itemize}
\item[] $s_0\mapsto \{ Cstatus_1 =in , Sender_1= reader\}$,

\item[] $s_1\mapsto\{Cstatus_1 = in, Sender_1= card \}$,

\item[] $s_2 \mapsto \{Cstatus_1 =out, Sender_1= card \}$,

\item[] $s_3 \mapsto \{Cstatus_1= out, Sender_1=reader \} \}$.
\end{itemize}

At the initial state, the chip card is inserted in the reader, and sending a
message must be done by the reader.  The protocol evolves by the action of four
events:
\begin{itemize}
\item \textit{Rsends:} the reader sends a message,
\item \textit{Csends:} the chip card sends a  message,
\item \textit{Eject:} the chip card is ejected,
\item \textit{Cinsert:} the chip card is inserted.
\end{itemize}

We assume that all the applications which use the protocol do not
request the transport of an infinite sequences of messages. This
is a fairness constraint which comes from the environment. It
ensures that the transmission of the messages between the card and
the reader terminates by the ejection of the card.

In $B$ event systems we proposed the following extension. The
fairness assumptions are written $(e\ \mathit{if} \ p)$
\cite{bellegarde01a}, where \textit{e} is the name of an event and
\textit{p} is a predicate characterizing the states in which
\textit{e} cannot be avoided when it is enabled infinitely often.
When an event must always be fair, the fairness assumption is only
the event name.

\begin{figure}
\begin{center}
\fontsize{6}{10}
\begin{boxedminipage}{10cm}
\begin{tabbing}
\textbf{MACHINE} teg1\\
\textbf{SETS} \= \\
\>$SENDER= \{card, reader\}$; $CARD-STATE =\{in,out\}$\\

\textbf{VARIABLES}\\
\>$Sender_1$, $Cstatus_1$\\

\textbf{INVARIANT}\\
\>$Sender_1 \in SENDER$ $\wedge$ $Cstatus_1 \in CARD-STATE$ \\

\textbf{INITIALISATION}\\
\>$Sender_1 := reader$ $\|$ $Cstatus_1:= in$ \\

\textbf{EVENTS}\\
\>\textbf{Rsends} $\BDef$ \= \textbf{SELECT} $Sender_1 = reader$
$\wedge$ $Cstatus_1= in$\\
\>\>\textbf{THEN} $Sender_1:= card  \textbf{END}$;\\

\>\textbf{Csends} $\BDef$ \textbf{SELECT} $Sender_1 = card$
$\wedge$
$Cstatus_1= in$ \\
\>\>\textbf{THEN} $Sender_1:= reader$  \textbf{END};\\

\>\textbf{Eject} $\BDef$ \textbf{SELECT} $Cstatus_1= in$ \\
\>\>\textbf{THEN} $Cstatus_1:= out$ \textbf{END};\\
. \>\textbf{Cinsert} $\BDef$ \textbf{SELECT} $Cstatus_1=out$
\\\>\> \textbf{THEN} $Sender_1:=reader$ $\|$ $Cstatus_1:=in$
  \textbf{END};\\
  \textbf{FAIRNESS}= \{Eject \};\\

\textbf{END}
\end{tabbing}
\end{boxedminipage}
\end{center}
\caption{Abstract B  specification of the protocol T=1 under
fairness assumptions} \label{facfigbspec1}
\end{figure}

\begin{figure}
\begin{center}
\includegraphics {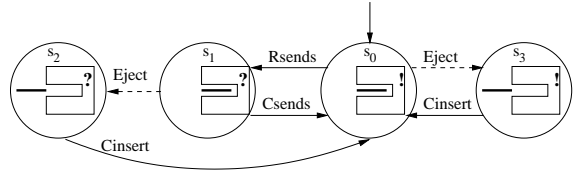}
\end{center}
\caption{Abstract fair transition system of the protocol T=1}
\label{fig1}
\end{figure}

The fairness assumption on the environment protocol is defined by
the clause \textit{FAIRNESS=\{Eject  \}} (see
Fig.~\ref{facfigbspec1}). In the abstract fair transition system
of the protocol, the fairness assumption is expressed by the set
$\enstranseq_1 = \{ F_{11} \}$ where $F_{11} = \{ s_0
\flsup{Eject} s_3, \ s_1 \flsup{Eject} s_2 \}$.

The set of states $\{s_0, s_1\}$ is defined by the following
expression : $Cstatus_1=in$. The expression $Cstatus_1 = out$
defines the set of states $\{s_2, s_3\}$. So, the verification of
a PLTL property $P$ on the transition system must be done under
this fairness assumption which is expressed by the PLTL formula:
\[\Box(\Box \Diamond   \ (Cstatus_1=in ) \ \Rightarrow \
\Diamond \  (Cstatus_1=out )).\]

This formula is a simplification of the formula obtained from the
general formula~\ref{eq-pltl-fairness} presented in
Section~\ref{sec:prelim} that would be:

$\Box(\Box \Diamond   \ (Cstatus_1=in  \ \et \ Sender_1= reader)
\vee (Cstatus_1=in  \ \et \ Sender_1= card)) \ \Rightarrow \
\Diamond \ ((Cstatus_1=out \ \et \ Sender_1= reader) \vee
(Cstatus_1= out  \ \et \ Sender_1= card))$.

\subsection{Refined Specification under Fairness Assumptions}
Figure~\ref{facfigbspec2} describes the refined $B$ event specification of the
protocol.  At this level of specification, we view the messages as a
sequence of
blocks ended by a last block (value \textit{lb}).  A block sent (value
\textit{bl}) is acknowledged by an acknowledgement block (value \textit{ackb}).
We call \textit{frame} these three types of exchanged information.

After a last block is sent by one of the devices, the other device
answers with a sequence of blocks ending by a last block unless
the card is ejected. These exchanges of messages alternate until
the card is ejected.

The fair transition system shown in Fig.~\ref{fig2} describes the
refined behavior of the protocol. The states of the transition
system are decorated by the value of variables $CardF_2$ and
$ReaderF_2$ which describe the type of the last frame sent
respectively by the chip card and the reader, $SenderF_2$ which
describes the device that will send the next frame, and
$Cstatus_2$ which does the same thing as the variable $Cstatus_1$.

The gluing invariant is a part of the invariant in
Fig.~\ref{facfigbspec2}:\\
$I_{12}= ((Cstatus_2 = Cstatus_1) \ \et$
\\ $((ReaderF_2= bl \ \Or \  ((CardF_2= ackb \ \Or \ CardF_2 = lb)
\ \et \  SenderF_2= reader ))\ \Leftrightarrow \ (Sender_1=reader)) \ \et$ \\
$ ((CardF_2= bl \ \Or \ ((ReaderF_2= ackb \ \Or \ ReaderF_2 = lb)
\ \et \ SenderF_2= card)) \ \Leftrightarrow \
(Sender_1= card)))$.

\mbox{}\par
With this gluing invariant, the states of the two transition
systems in Fig.~\ref{fig1} and Fig.~\ref{fig2} are glued as follows:
\begin{itemize}
\item $r_0, r_2, r_3, r_4, r_{10}$ are glued with the state $s_0$,
\item $r_5, r_7, r_8,  r_9,   r_{12}$ are glued with the state  $s_1$,
\item $r_6, r_{13}$ are glued with the state  $s_2$,
\item $r_1, r_{11}$ are glued with the state  $s_3$.
\end{itemize}
From these four equivalence classes (see
Section.~\ref{subsec:part-by-ref}) we construct the four parts
described in Fig.~\ref{mod1}, Fig.~\ref{mod2}, Fig.~\ref{mod6} and
Fig.~\ref{mod4} (see the appendix).

The state labelling function, called $L_2$ is the following:$L_2 =\{$
\begin{itemize}
\item[] $r_0 \mapsto \{ Cstatus_2 =in ,\  SenderF_2= reader,\
ReaderF_2= lb, \ CardF_2= lb\}$,

\item[] $r_1\mapsto \{Cstatus_2 = out, \ SenderF_2= reader,   \ 
ReaderF_2= lb, \ CardF_2= lb\}$,

\item[] $r_2\mapsto \{Cstatus_2 = in, \ SenderF_2=  card, \ ReaderF_2= bl,
\ CardF_2= lb\}$,

\item[] $\ \ \ \ldots\ \ \ \}$.
\end{itemize}

The old events $Rsends$ and $Csends$ terminate the
emission of a message by sending the last block. We have
reinforced the guard of the event $Eject$ in order to forbid the
ejection of the chip card during the transmission of a message. We
have introduced four new events to take the transmission of blocks
and acknowledgement into account:
\begin{itemize}
\item \textit{Rblocksends:} the reader sends a block,
\item \textit{Cblocksends:} the card sends a block,
\item \textit{Racksends:} the reader sends an acknowledgement,
\item \textit{Cacksends:} the card sends an acknowledgement.
\end{itemize}

The fairness assumptions in the refined level are defined by the
declaration set \textit{FAIRNESS=\{Eject, Csends if
$(CardF_2=bl)$, Rsends if $(ReaderF_2 =bl)$\}}. The fairness
assumption $h_1 \egdef  Eject$  reformulates the abstract fairness
assumption. It indicates that the end of the transmission is
unavoidable. The fairness assumptions $h_2\egdef Csends \
\mathit{if} \ (CardF_2=bl)$ and $h_3 \egdef Rsends \ \mathit{if} \
(ReaderF_2 =bl)$ are new fairness assumptions which express that
the messages sent by the chip card or the reader contain a finite
number of blocks. They allow to go out of livelocks as shown in
Fig.~\ref{fig2}. These assumptions must be satisfied by all the
possible environments of the protocol.

The verification of a PLTL property $P$ on the transition system
must be done under these fairness assumptions, expressed by the
following PLTL formulae:
\\
\begin{itemize}
\item $f_1'\egdef \Box$($\Box$ $\Diamond$ ($Cstatus_2 = in  \ \et
\ ((SenderF_2 = reader  \ \et \ CardF_2= lb) \ \vee \ (SenderF_2 =
card \ \et \ ReaderF_2 = lb)))$ $\Rightarrow$ $\Diamond$
$Cstatus_2= out )$ expresses $h_1$,\\

\item $f_{21} \egdef  \Box$($\Box$ $\Diamond$($SenderF_2= card$
$\et$ $CardF_2= bl)$ $\Rightarrow$ $\Diamond$ $CardF_2=lb)$
expresses $h_2$,\\

\item $f_{22} \egdef  \Box$($\Box$ $\Diamond$ ($SenderF_2= reader$
$\et$ $ReaderF_2$= bl) $\Rightarrow$ $\Diamond$ $ReaderF_2 = lb)$
  expresses $h_3$.
\end{itemize}

The fairness assumptions  are defined  by the sets of  fair
transitions (represented in  Fig.~\ref{fig2} by the dashed arrows)
$\enstranseq_2 = \{F_{1}', F_{21}, F_{22} \}$ where :

\begin{itemize}
\item $F_{1}' \egdef \{r_{0} \flsup{Eject} r_{1}, \ r_{10}
\flsup{Eject} r_{11}, \  r_{12} \flsup{Eject} r_{13}, r_5
\flsup{Eject} r_6 \}$  which formalizes $h_1$, \item $F_{21}
\egdef \{r_8 \flsup{Csends} r_{10}\}$ which formalizes $h_2$,
\item $F_{22} \egdef \{r_3 \flsup{Rsends} r_5\}$ which formalizes
$h_3$.
\end{itemize}

\begin{figure}
\begin{center}
\fontsize{6}{10}
\begin{boxedminipage}{15cm}

\begin{tabbing}
\textbf{REFINEMENT} teg1ref \textbf{REFINES} teg1\\

\textbf{SETS}\= \\
\> $FRAME = \{bl,lb, ackb\}$\\
\textbf{VARIABLES}\\
\>$SenderF_2$, $Cstatus_2$, $CardF_2$, $ReaderF_2$ \\

\textbf{INVARIANT}\\
\>$SenderF_2 \in SENDER $ $\wedge$ $CardF_2 \in FRAME$ $\et$
$ReaderF_2  \in FRAME$ $\et$ $Cstatus_2 = Cstatus_1$
$\et$ \\
\>($ReaderF_2= bl$ $\Or$ (($CardF_2= ackb$ $\Or$ $CardF_2 = lb$)
$\et$ $SenderF_2= reader$)) $\Leftrightarrow$ ($Sender_1= reader$)
$\et$
  \\ \>($CardF_2= bl$ $\Or$ (($ReaderF_2= ackb$ $\Or$ $ReaderF_2 =
lb$) $\et$ $SenderF_2= card$)) $\Leftrightarrow$ ($Sender_1= card$)\\

\textbf{INITIALISATION}\\
\>$SenderF_2:= reader$ $\|$ $Cstatus_2:= in$ $\|$  $CardF_2:= lb$
$\|$ $ReaderF_2:= lb$ \\

\textbf{EVENTS}\\
\>\textbf{Rsends} $\BDef$ \= \textbf{SELECT} ($SenderF_2 = reader$
$\wedge$ $Cstatus_2= in$ \\ \>\> $\et$ ($CardF_2= ackb$ $\Or$ $CardF_2= lb$))\\
\>\>\textbf{THEN} $SenderF_2:= card$ $\|$  $ReaderF_2:= lb$  \textbf{END};\\

\>\textbf{Csends} $\BDef$ \textbf{SELECT} ($SenderF_2 = card$
$\wedge$ $Cstatus_2= in$  \\ \>\> $\et$ ($ReaderF_2= ackb$ $\Or$
$ReaderF_2= lb$)) \\
\>\>\textbf{THEN} $SenderF_2:= reader$ $\|$  $CardF_2:= lb$  \textbf{END};\\

\>\textbf{Eject} $\BDef$ \textbf{SELECT} (($SenderF_2=card$ $\et$
$ReaderF_2=lb$) $\Or$ ($SenderF_2= reader$ $\et$ $CardF_2= lb$))
$\et$\\
\>\>$Cstatus_2=in$ \\ \>\>  \textbf{THEN} $Cstatus_2:= out$ \textbf{END};\\

\>\textbf{Cinsert} $\BDef$ \textbf{SELECT} $Cstatus_2=out$ \\\>\>
\textbf{THEN} $SenderF_2:=reader$ $\|$ $Cstatus_2:=in \ \| \ CardF_2 := lb \ \| \ ReaderF_2 := lb$  \textbf{END};\\

\>\textbf{Rblocksends} $\BDef$ \textbf{SELECT} ($SenderF_2 =
reader$ $\wedge$ $Cstatus_2= in$
  $\et$ ($CardF_2= ackb$ $\Or$ $CardF_2= lb$) )\\
\>\>\textbf{THEN} $SenderF_2:= card$ $\|$  $ReaderF_2:= bl$   \textbf{END};\\

\>\textbf{Cblocksends} $\BDef$ \textbf{SELECT} ($SenderF_2 = card$
$\wedge$ $Cstatus_2= in$
  $\et$ ($ReaderF_2= ackb$ $\Or$ $ReaderF_2= lb$) )\\
\>\>\textbf{THEN} $SenderF_2:= reader$ $\|$  $CardF_2:= bl$  \textbf{END};\\

\>\textbf{Racksends} $\BDef$ \textbf{SELECT} ($SenderF_2 = reader$
$\wedge$ $Cstatus_2= in$
  $\et$  $CardF_2=bl$)\\
\>\>\textbf{THEN} $SenderF_2:= card$ $\|$  $ReaderF_2:= ackb$ \textbf{END};\\

\>\textbf{Cacksends} $\BDef$ \textbf{SELECT} ($SenderF_2 = card$
$\wedge$ $Cstatus_2= in$
  $\et$  $ReaderF_2=bl$)\\
\>\>\textbf{THEN} $SenderF_2:= reader$ $\|$  $CardF_2:= ackb$ \textbf{END};\\
\textit{\textbf{FAIRNESS}=\{Eject , Csends if $(CardF_2=bl)$,
Rsends if
$(ReaderF_2 =bl)$\}};\\

\textbf{END}
\end{tabbing}
\end{boxedminipage}
\end{center}
\caption{Refined B  specification of the protocol T=1 under
fairness assumptions} \label{facfigbspec2}
\end{figure}

\begin{figure}
\begin{center}
\includegraphics {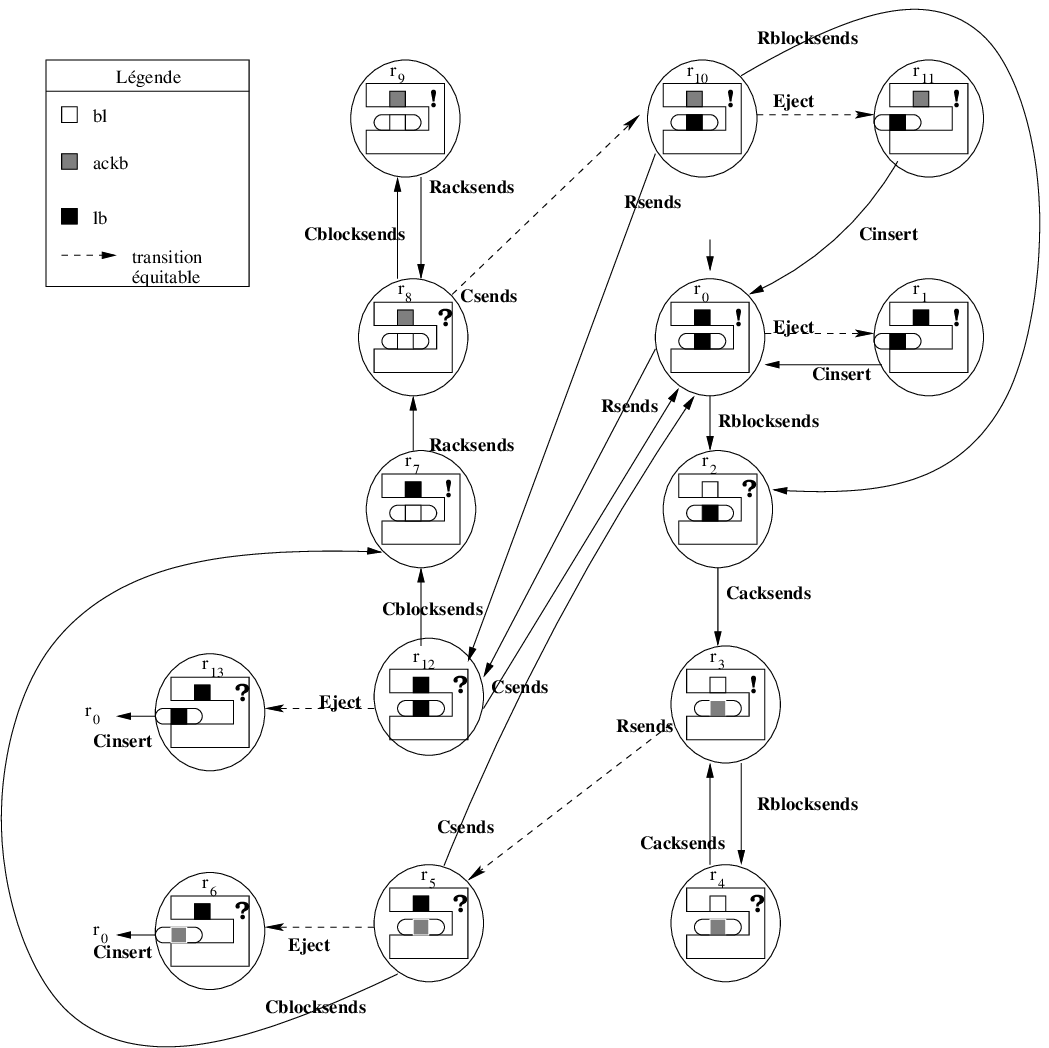}
\end{center}
\caption{Refined fair transition system of the protocol T=1}
\label{fig2}
\end{figure}

\subsection{Example of the verification by parts}
\label{sec-expl-verif-part} In this section, we present an
application of partitioned model checking on the example of the
protocol T=1. We use the tool SPIN \cite{Holzmann91} for the
verification by model checking.  The test consists in splitting
the transition system of the protocol T=1, and in choosing a PLTL
property $P$ verifiable by parts. Then, we verify $P$ under
fairness assumptions $f$ on the global system, and on each part.
Our goal is to illustrate that the property $f \Rightarrow P$ is
verifiable in a partitioned way.

Figures~\ref{mod1} to \ref{mod4} (see the appendix) represent the
parts which are obtained by splitting the global transition system in
Fig.~\ref{fig2} at the refined level from Definition~\ref{def:module}.

Let us verify $P \egdef \Box (CardF_2 = bl  \Rightarrow \Diamond
(CardF_2 = lb \wedge SenderF_2 = reader ))$ which expresses that
when a card sends a block, then it will inevitably send  a last
block. We verify $P$ by refinement based parts under the fairness
assumptions $f_1', f_{21}, f_{22}$. The PLTL formula which
expresses the fairness assumptions is $f \egdef f_1' \wedge f_{21}
\wedge f_{22}$. The property to be verified is $Q \egdef f
\Rightarrow P$.

First, we checked that $Q$ is satisfied on the global system of the
T=1 protocol by the tool SPIN. Second, we verified $Q$ on each part in
the following way.  Before checking $Q$ on the part $s_0$ such as
described in the Fig.~\ref{mod1}, we simplified $Q$ because the part
$s_0$ is only concerned with the fairness assumptions $f_{22}$.  This
part does not contain a cycle, which is forbidden by the fairness
assumptions $f_1'$ or $f_{21}$, therefore these assumptions are
useless to check $Q$ in this part.  Thus we simplified $Q$ in $Q_1
\egdef f_{22 } \Rightarrow P$.  We used SPIN to check that $Q_1$ is
satisfied on the part $s_0$, therefore $Q$ is satisfied on $s_0$.  The
part $s_1$ in Fig.~\ref{mod2} is concerned with the fairness
assumptions $f_{21}$.  So we simplified $Q$ in $Q_2 \egdef f_{21 }
\Rightarrow P$.  The property $Q_2$ is satisfied on the part $s_1$,
therefore the property $Q$ is satisfied on $s_1$.  The parts $s_2$ in
Fig.~\ref{mod6} and $s_3$ in Fig.~\ref{mod4} do not contain cycles
forbidden by the fairness assumptions, all their executions are fair.
Thus they are not concerned with fairness assumptions.  So, we
simplified $Q$ in $Q_3 \egdef P$.  The property $Q_3$ is satisfied on
the parts $s_2$ and $s_3$, therefore the property $Q$ is satisfied
also on these parts.  Accordingly to Theorem~\ref{th-verifpart}, the
property $Q$ is verifiable by refinement based parts (see
Definition~\ref{def:module}).  Since all the parts satisfy $Q$, the
global system of the protocol T=1 satisfies $Q$.

\subsection{Counter example if the condition of decomposition by
refinement does not holds}
In this section, we give an example that shows that the approach of
verifying by parts PLTL property under fairness assumptions is not
correct, when the condition (see Definition~\ref{def:module}) of the
decomposition by refinement is not satisfied.  We propose a
decomposition of the system so that the executions of the parts cannot
leave the $\tau$-cycles and fair exiting cycles, contained in the
parts resulting from the decomposition.  In this example, we note
$(r_i \flsup{a_i} \ldots \flsup{a_{i+n}} r_i)^*$ a finite fragment of
an execution which runs around a cycle finitely many times.  We note
$(r_i \flsup{a_i} \ldots \flsup{a_{i+n}} r_i)^{\omega}$ an infinite
fragment of an execution which runs around a cycle infinitely many
times.

Figures~\ref{mod7} and~\ref{mod8} represent the parts $s_1'$ and
$s_2'$ obtained by splitting the global system of the protocol
T=1, without using the refinement.  Notice that, we have not used
a refinement based partitioning but the partition of the
transitions of the system.

Let us verify  the PLTL property $P'  \egdef \Box (SenderF_2 =
reader \ \et \ ReaderF_2 = lb \ \et \ Cstatus_2 = in \Rightarrow
\Diamond (ReaderF_2 = lb \et Cstatus_2 = out)) $, which expresses
that if the reader sends a last block then the card eventually
will eject. We verify $P'$ under the fairness assumptions  $f =
f_1 ' \ \et \ f_{21 } \ \et \ f_{22}$. Notice that $P'$ is
verifiable by  parts, because the B\"{u}chi automaton of $\neg P'$
belongs to the class $\ABmod$.  The property to be verified by
parts is $Q' \egdef f \Rightarrow P' $.

We verified $Q'$ on the global system of the protocol T=1 using
SPIN. Then, we verified $Q'$ on each part, $s_1'$ and $s_2'$.  The
results are that $Q'$ is violated on the global system and
satisfied on the parts.  In the part $s_2'$, the executions are
not fair. Since $f$ is false on $s_2'$  then $Q' \egdef f
\Rightarrow P'$ is satisfied on  $s_2'$. $Q'$ is satisfied on the
part $s_1'$ because the only computations of this part are those
which reach infinitely many times the states $r_0$ and $r_1$ in
Fig.~\ref{mod7}.  These executions satisfy $Q'$.  The other
executions of $s_1'$ are not fair, therefore they satisfy $Q'$.
Consequently, the property $Q'$ is not verifiable by parts
although the B\"uchi automaton of $\neg P'$ belongs to the class
$\ABmod$, i.e. is verifiable by parts. We recall that a PLTL
property $P$ under fairness assumptions $f \Rightarrow P$ is
verifiable by parts under the following condition : "if the
property is not satisfied on the global system, then there is a
part which violates the property".

With such a decomposition of the global system, the method of
verifying by parts is not correct because  the parts do not
contain  fragments of some computations of the global system. For
example, the computation 
\begin{itemize}
\item[] $\sigma_1 \egdef (r_0 \flsup{Rblocksends}
r_2 \flsup{Cacksends}(r_3\flsup{Rblocksends} r_4 \flsup{Cacksends}
r_3) ^* \flsup{Rsends} r_5 \flsup{Eject}r_6 \flsup{Cinsert}
r_0)^{\omega}$
\end{itemize}
in Fig.~\ref{fig2}, does not have fragments in the parts $s_1'$ and
$s_2'$.  Indeed the executions
\begin{itemize}
\item[] $\sigma_2 \egdef r_0
\flsup{Rblocksends} r_2 \flsup{Cacksends}(r_3\flsup{Rblocksends} r_4
\flsup{Cacksends} r_3)^{\omega}$
\end{itemize}
in the part $s_1'$ and 
\begin{itemize}
\item[] $\sigma_3
\egdef (r_0 \flsup{Rsends} r_{12} \flsup{Csends} r_0)^{\omega}$
\end{itemize}
in the part $s_2'$, are not extensions of fragments of the execution
$\sigma_1$.  As $\sigma_2$ and $\sigma_3$ are not fair, then they
satisfy the property $Q'$.  The method fails because there are
computations of the global transition system which are not verified in
the parts.  This because their fragments do not exist in the
computations of the parts.  On the other hand the decomposition by
refinement ensures that all the fragments of all the computations of
the global system are in the parts.  For example, the computation
$\sigma_1$ is broken into two fragments which we find in the
computations of the parts obtained by refinement.  The first fragment
of $\sigma_1$ is
\begin{itemize}
\item[] $\varphi_1 \egdef r_0 \flsup{Rblocksends} r_2
\flsup{Cacksends} (r_3\flsup{Rblocksends} r_4 \flsup{Cacksends} r_3)*
\flsup{Rsends } r_5 \flsup{Eject}r_6$ 
\end{itemize}
which is a prefix of the computation 
\begin{itemize}
\item[] $\sigma_4 \egdef r_0 \flsup{\textit{\scriptsize
Rblocksends}} r_2 \flsup{\textit{\scriptsize Cacksends}}
r_3\flsup{\textit{\scriptsize Rblocksends}} (r_4
\flsup{\textit{\scriptsize Cacksends}}
r_3\flsup{\textit{\scriptsize Rblocksends}} r_4)^*
\flsup{\textit{\scriptsize Rsends}} r_5 \flsup{\textit{\scriptsize
Eject}} (r_6 \flsup{\textit{\scriptsize skip}}r_6)^{\omega}$
\end{itemize}
in the part $s_0$ in Fig.~\ref{mod1}.  We find also the second
fragment 
\begin{itemize}
\item[] $\varphi_2 \egdef r_6 \flsup{Cinsert} r_{0}$
\end{itemize}
as a prefix of the computation
\begin{itemize}
\item[] $\sigma_5 \egdef r_6 \flsup{Cinsert } r_{0}
\flsup{Eject}(r_1 \flsup{skip}r_1)^{\omega }$
\end{itemize}
in the part $s_2$ in Fig.~\ref{mod6}.  Let us note that $\sigma_1$ is
the concatenation of these two fragments.  The two executions
$\sigma_4$ and $\sigma_5$ which contain the fragments $\varphi_1$ and
$\varphi_2$ are fair, therefore they satisfy $f$.  Since $\sigma_4$
and $\sigma_5$ satisfy also $Q \egdef f \Rightarrow P$ as verified
previously --~$Q$ is the property verified by parts with a decomposition
by refinement in Section~\ref{sec-expl-verif-part}~-- therefore they
satisfy $P$.  Since $P$ is verifiable by parts, then the proof that
$P$ is satisfied on $\sigma_4$ and $\sigma_5$ is a proof that $P$ is
satisfied on $\sigma_1$.

\section{Performance of the approach of the verification by part}
\label{sec:expResults} In this section, we show the experimental
results of the approach of the verification by refinement based
parts. The expected performance is the capacity of this approach
to verify  a large set of different type of properties at the
refined level. So, we verify  two examples of applications. The
first example is the protocol T=1, the second one is the car
wind-screen wipers system.  For each example we choose different
PLTL properties to verify.  The properties must express the
general behaviors of the systems.

At the end of this section, we compare our approach with the
Pnueli's approach~\cite{kesten98,kesten01}, that is another
verification approach under fairness assumptions.

The protocol T=1 at the refined level is described in the last
section.  We propose to verify by parts the following PLTL properties
which express the main behaviors of the protocol :

\begin{itemize}

\item These properties express that always when a device send a
block then it will inevitably send a last block:

 $P_1 \egdef \Box(CardF_2= bl \Rightarrow \Diamond ( CardF_2
= lb $)).  $P_2 \egdef   \Box(ReaderF_2= bl \Rightarrow \Diamond (
ReaderF_2 = lb $)).

\item These properties express that always when a device send a
block then the other device will inevitably send  an
acknowledgment block:

$P_3 \egdef  \Box(CardF_2= bl  \Rightarrow \Diamond ( ReaderF_2 =
ackb $)).   $P_4  \egdef  \Box(ReaderF_2= bl \Rightarrow \Diamond
( CardF_2 = ackb $)).

\item This property expresses that when the card sends a block and
the reader sends an acknowledgment block,  the card and the reader
will respectively send an acknowledgment block and a block:

$P_5 \egdef  \Box(CardF_2= bl \ \et \ ReaderF_2= ackb \Rightarrow
\Diamond ( CardF_2 = ackb \ \et \ ReaderF_2= bl$)).

\item This property expresses the alternation of sending of the
messages between the card and the reader:

$P_6 \egdef  \Box(SenderF_2= card  \Rightarrow \Diamond (
SenderF_2= reader $)).

\end{itemize}

The second application verified is the car wind-screen wipers
system. At the refined level the system is composed of a control
level, a rain sensor and two (left and right) wind-screen wipers.
The control level can select the mode - automatic or manual- of
the wiper system. The left and the right wipers have the same
behavior. The rain sensor can detect the rain amount (no rain,
small rain, strong rain). We have verified six properties on this
application. The results are shown in the following section.

\subsection{Results of the verification}

The following table indicates the results.  We give the number of
properties to verify, how many are globally true, how many
globally false, and how many have been successfully verified by
refinement based parts.

\bigskip
\begin{center}
\begin{small}
\begin{tabular}{|c|c|c|c|c|}
     \hline
     Example & Properties & Globally true & Globally false &
     Verified by parts  \tabularnewline
     \hline\hline
     protocol T=1 & 6 & 5 & 1 & 4  \tabularnewline
     \hline
     wind-screen wipers& 6 & 6 & 0 & 4  \tabularnewline
     \hline
\end{tabular}
\end{small}
\end{center}
\bigskip

These results show that we have successfully verified four
properties by parts amongst the six that were expressed, on the
protocol T=1 system and the car wind-screen wipers system. In the
case of the protocol T=1, the verification  failed for properties
$P_5$ and $P_6$. As $P_5$ is globally false, then there is at
least a part on which it is false, that is  $s_0$. In contrast the
property $P_6$ is globally true and our method failed to prove it
 by refinement based parts (it is false on parts $s_0$ and
$s_1$). This is due to the fact that $P_6$ is not a \emph{new}
property. It expresses an abstract behavior of the system, and
should have been verified by parts at a former level of
refinement. Properties $P_1$, $P_2$, $P_3$ and $P_4$ express new
behaviors of the system at the refined level, and as expected,
they have been successfully verified by parts.

We obtained the same results for the verification of the car
wind-screen wipers system. The approach of the verification by
parts fails in verifying two properties that have been verified at
the abstract level. However, the approach succeeds in verifying
properties that express the new behaviors of the system at the
refined level.

These results show that as  it would be interesting to study the
relation between our approach of verification, the new properties
at the refined level, and the abstract properties in order to
characterize the properties for which a verification by parts is
suited.

\subsection{Comparison with Pnueli's approach of verification}
An interesting method was proposed in~\cite{kesten98,kesten01},
this is a symbolic model checking of PLTL properties under
fairness assumptions.  This approach removes the fairness
assumptions from the formula to verify.  It deals with a fairness
assumptions at the algorithmic level instead of specifying them as
of a part of the formula to be verified. Fairness assumptions are
expressed as B\"{u}chi (for weak fairness) and Street (for strong
fairness) automata acceptance conditions. So, this algorithm
verifies the property $P$ instead of verifying $f \Rightarrow P$.
The verification consists of the emptiness checking which is
implemented  with BDD.

This approach treats the problem of the combinatorial explosion of
model checking by simplifying the formula to verify under fairness
approach.  In our case we treats the problem of the combinatorial
explosion by the following way.   We partition the transition
system and we verify a property separately on each part. Another
difference is that we adapted the automata algorithm of Vardi and
Wolper \cite{Vardi86} instead of a symbolic algorithm. However, we
can combine our approach with Pnueli's approach. So in order to
verify a property $P$ under fairness assumptions $f$, we verify
the simplified formula $P$ by the partitioning way. So we will use
Pnueli's approach to express fairness assumptions by Street and
B\"{u}chi automata conditions on each transition system of the
parts. Then we will exploit our approach to split the global
transition system using the fair refinement relation. Finally we
will verify $P$ on each part using symbolic model checking instead
of verifying $f\Rightarrow P$ on the global system.

\section{Conclusion and Future Works}
\label{sec:conlu} In this paper, we  extend the partitioned model
checking technique presented in \cite{JulliandTSI01} to handle the
fairness constraints of the system environment.  Our goal is to
verify the PLTL properties under fairness assumptions by part.
When the fairness constraints of the environment are expressed by
fairness assumptions, the verification by model checking of a
partitioned property $P$ under fairness assumptions $f$ supposes
verifying by model checking the new property $Q \egdef f \Impl P$
on the transition system. However, the property $Q$ does not
necessarily belong to the class of properties verifiable by part.

Our contribution in this paper is to prove that the split of the
transition system into parts, using the fair refinement relation,
makes the property $Q$ verifiable by refinement based parts when
$P$ is verifiable by part. The use of the fair refinement to split
a transition system allows us to obtain refinement based parts
which contains computations.  This is a sufficient condition to
verify by refinement based parts the property $Q \egdef f \Impl
P$. To handle the fairness constraints, we have proposed to use a
fair transition system to model a reactive system and its fairness
environment. This framework is a transition system which contains
only computations.

The complexity of the refinement verification is linear in the
size of the refined system,
 because it necessitates only an enumeration of the refined model when the
gluing relation is a function. Therefore the verification by parts
is interesting because the additive decomposition of a system
comes for free from the refinement verification.

In the future, we plan to implement the partitioned model checking
technique so that we can evaluate its performance on industrial
applications. As we saw in the example of the protocol T=1, we can
simplify the fairness assumptions on each part because all the
assumptions do not concern every part. So, the simplification
process needs to be further studied. We plan also to combine the
approach of model checking by part to the approach
\cite{chouali&03} of model checking under fairness assumptions
which exploits the fair refinement in order to reduce the size of
the formula to be verified under fairness assumptions.

Also, we must give simplification rules to translate fairness
assumptions, as expressed in the event systems, into PLTL
formulae, as used by the usual model checking algorithms.  We plan
to study a variation of this method which does not requires the
condition (b) in Definition \ref{def-FTS}.  In this case the
fairness assumptions are not expressible in PLTL, but in a logic
of actions as the $\mu$-calculus \cite{Clarke99}.

\bibliographystyle{alpha}
\bibliography{biblio-acm-tecs}

\newpage
\renewcommand{\floatpagefraction}{.10}
        \renewcommand{\textfraction}{.0}

\appendix
\section *{Appendix}
\setcounter{section}{1}

\begin{figure}[ht]
\begin{center}
\includegraphics {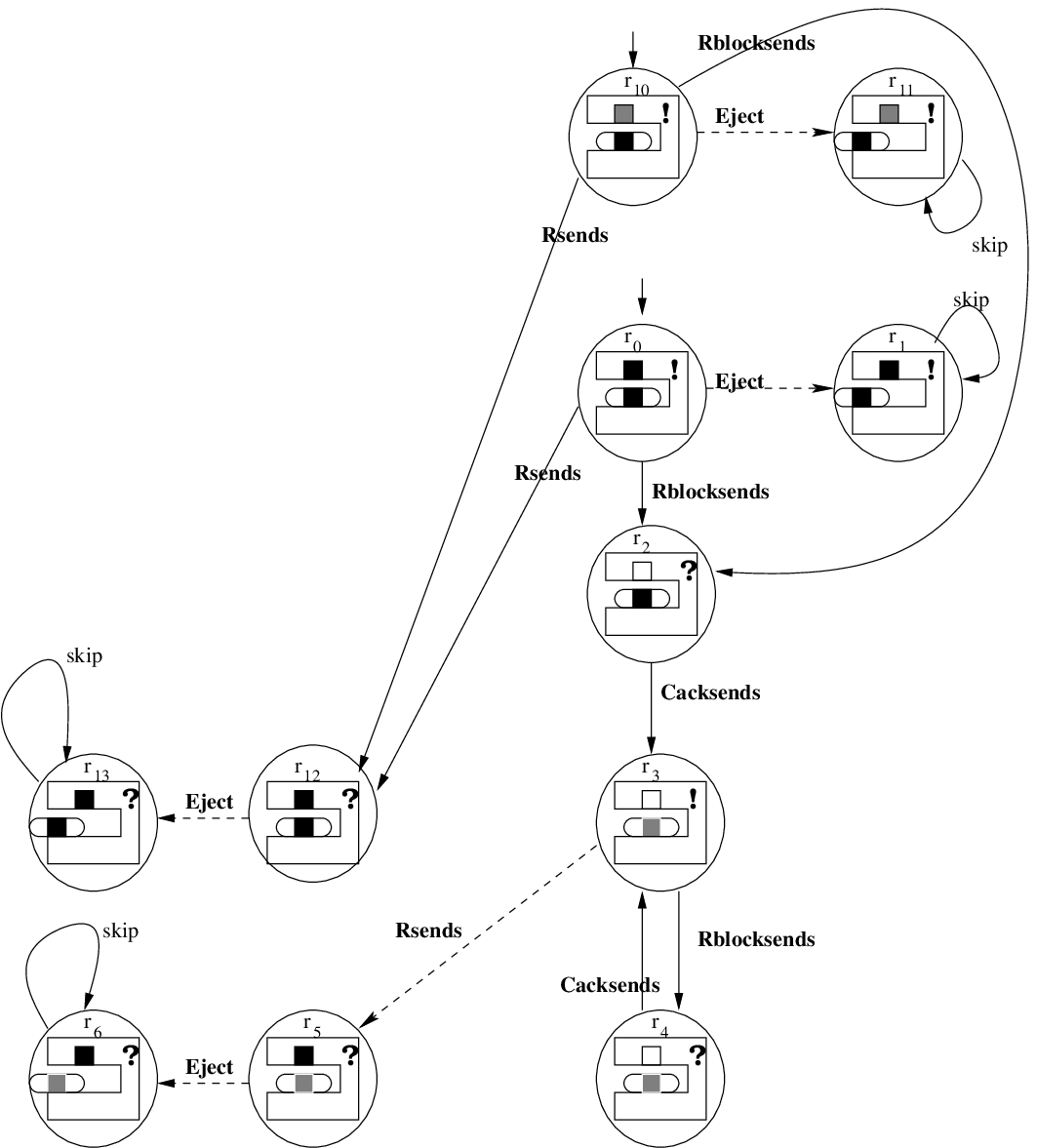}
\end{center}
\caption{The part $s_0$ of the protocol T=1} \label{mod1}
\end{figure}

\begin{figure}
\begin{center}
\includegraphics {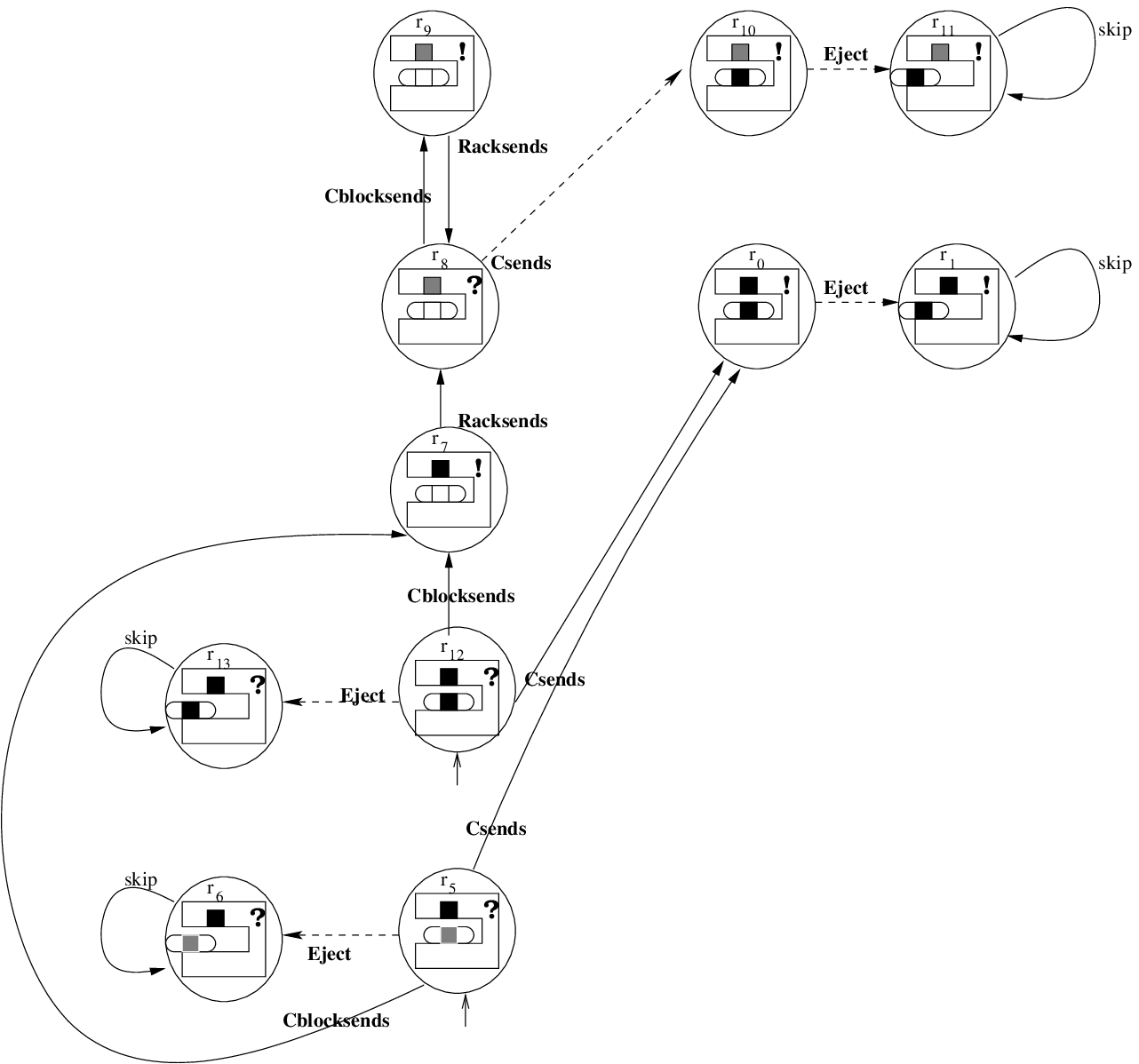}
\end{center}
\caption{The part $s_1$ of the protocol T=1} \label{mod2}
\end{figure}

\begin{figure}
\begin{center}
\includegraphics {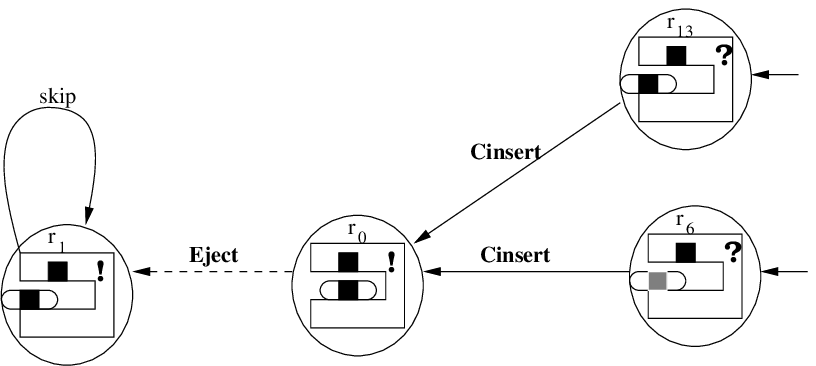}
\end{center}
\caption{The part $s_2$ of the protocol T=1} \label{mod6}
\end{figure}

\newpage

\begin{figure}
\begin{center}
\includegraphics {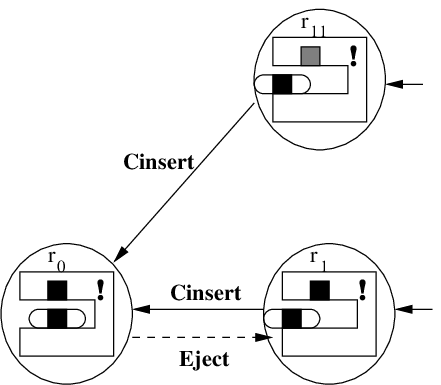}
\end{center}
\caption{The part $s_3$ of the protocol T=1} \label{mod4}
\end{figure}

\begin{figure}[t]
\begin{center}
\includegraphics {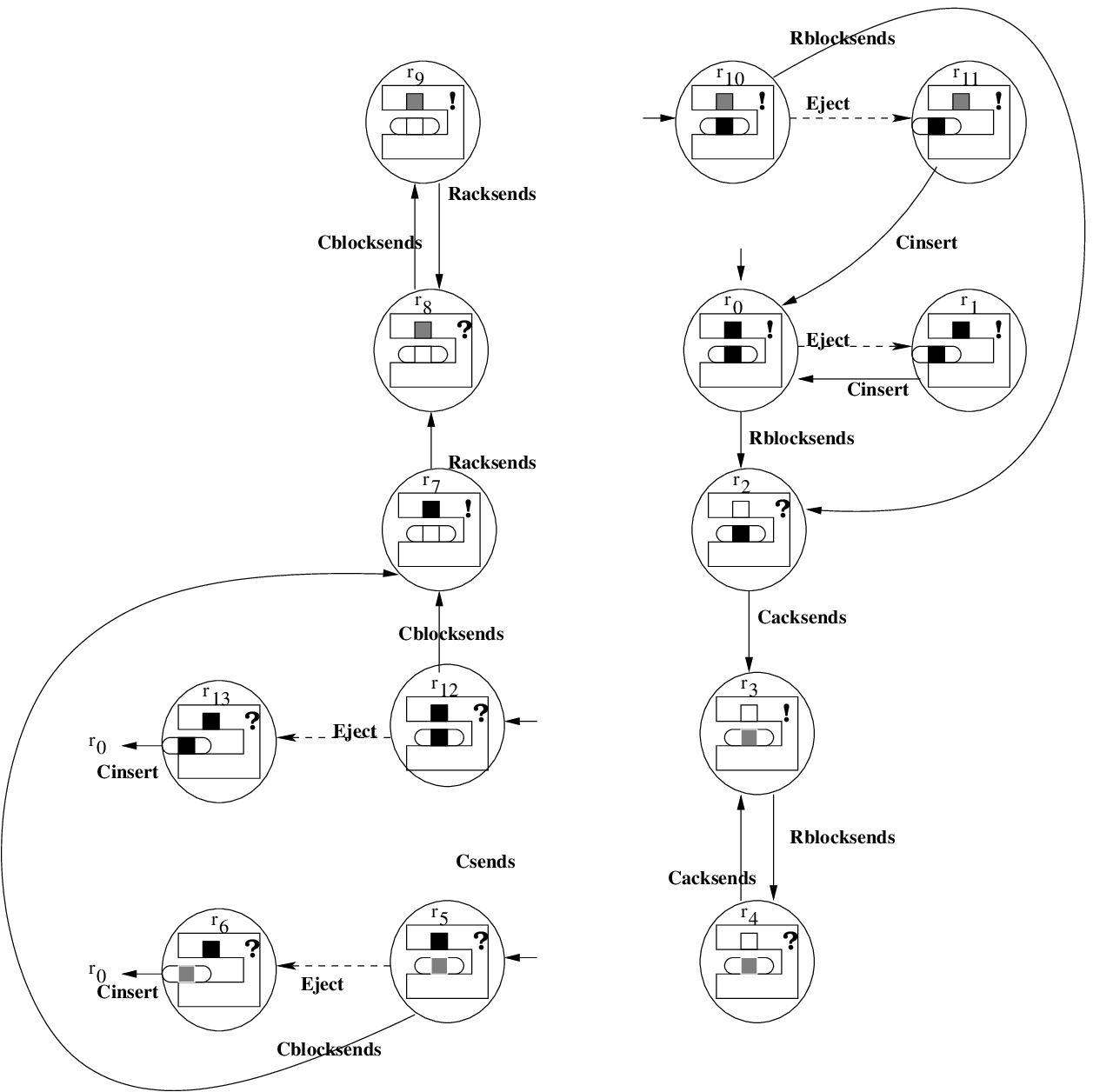}
\end{center}
\caption{The part $s_1'$ of the protocol T=1} \label{mod7}
\end{figure}

\newpage

\begin{figure}[t]
\begin{center}
\includegraphics {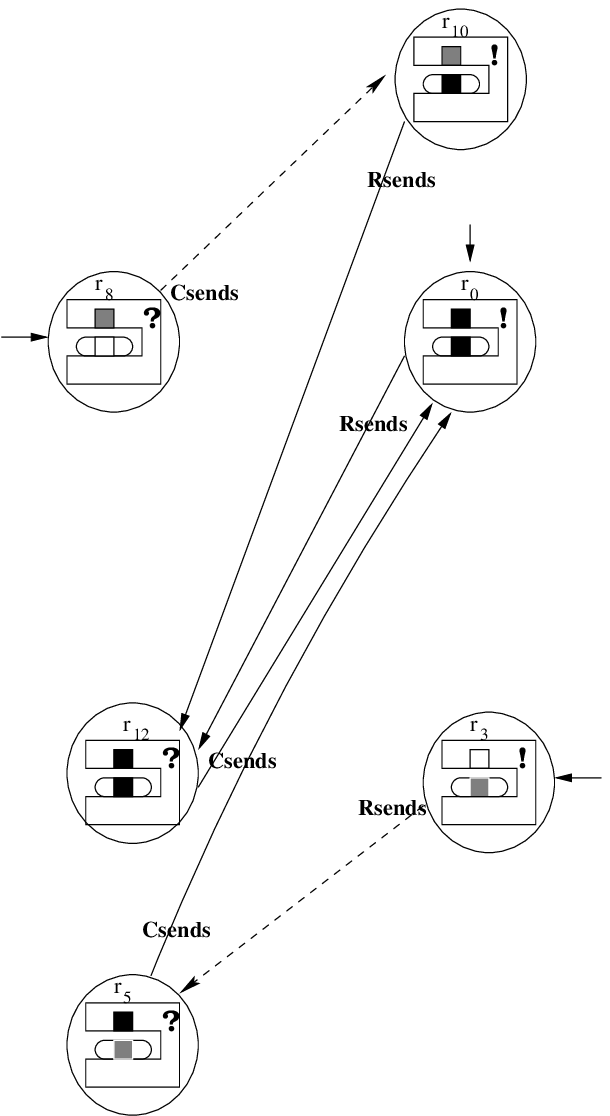}
\end{center}
\caption{The part $s_2'$ of the protocol T=1} \label{mod8}
\end{figure}

\end{document}